\documentclass[Royal,sagev]{sagej}   

\usepackage{amsthm,amssymb,amsmath,amsxtra,array,amsfonts,dsfont,enumerate,graphicx,multirow,moreverb,natbib,setspace,subfig}

\usepackage[colorlinks,bookmarksopen,bookmarksnumbered,citecolor=black,urlcolor=black]{hyperref}

\usepackage{rotating}

\theoremstyle{plain}

\theoremstyle{definition}

\theoremstyle{remark}

\begin{document}

\runninghead{Li, Lee, Port, Robinson}

\title{The Impact of Unmeasured Within- and Between-Cluster Confounding on the Bias of Effect Estimators from Fixed Effects, Mixed Effects and Instrumental Variable Models}

\author{Yun Li\affilnum{1, 2, 3}, Yoonseok Lee\affilnum{4}, Friedrich K Port\affilnum{3} and Bruce M Robinson\affilnum{3} }

\affiliation{\affilnum{1}Department of Biostatistics, Epidemiology and Informatics, University of Pennsylvania, Philadelphia, PA.\\
\affilnum{2} Department of Biostatistics, University of Michigan, Ann Arbor, MI. \\
\affilnum{3} Arbor Research Collaborative for Health, Ann Arbor, MI. \\
\affilnum{4} Department of Economics and Center for Policy Research, Syracuse University, Syracuse, NY.}

\corrauth{Yun Li}

\email{yun.li@pennmedicine.upenn.edu}

\begin{abstract}
Unmeasured confounding almost always exists in observational studies and can bias estimates of exposure effects. Instrumental variable methods are popular choices in combating unmeasured confounding to obtain less biased effect estimates. However, we demonstrate that alternative methods may give less biased estimates depending on the nature of unmeasured confounding. Treatment preferences of clusters (e.g., physician practices) are the most frequently used instruments in instrumental variable analyses (IVA). These preference-based IVAs are usually conducted on data clustered by region, hospital/facility, or physician, where unmeasured confounding often occurs within or between clusters. We aim to quantify the impact of unmeasured confounding on the bias of effect estimators in IVA, as well as several common alternative methods including ordinary least squares regression, linear mixed models (LMM) and fixed effect models (FE) to study the effect of a continuous exposure (e.g., treatment dose) on a continuous outcome. We derive closed-form expressions of asymptotic bias of estimators from these four methods in the presence of unmeasured within- and/or between-cluster confounders. Simulations demonstrate that the asymptotic bias formulae well approximate bias in finite samples for all methods. The bias formulae show that IVAs can provide consistent estimates when unmeasured within-cluster confounding exists, but not when between-cluster confounding exists \cite{LiY}. On the other hand, FEs and LMMs can provide consistent estimates when unmeasured between-cluster confounding exits, but not for within-cluster confounding. Whether IVAs are advantageous in reducing bias over FEs and LMMs depends on the extent of unmeasured within-cluster confounding relative to between-cluster confounding. Furthermore, the impact of unmeasured between-cluster confounding on IVA estimates is larger than the impact of unmeasured within-cluster confounding on FE and LMM estimates. We illustrate the use of these methods in estimating the effect of erythropoiesis stimulating agents on hemoglobin levels. Our findings provide guidance for choosing appropriate methods to combat the dominant types of unmeasured confounders and help interpret statistical results in the context of unmeasured confounding.
\end{abstract}


\maketitle

\section{Introduction}

Unmeasured confounding almost always exists in observational studies and can often bias the estimates of exposure effects. For example, patients' co-morbid conditions are common confounders for treatment effects. We may know of patients' co-morbid conditions but we usually do not have detailed information on the levels of disease severity, which can lead to unmeasured confounding and subsequent bias. Observational studies are often conducted with individuals clustered by region, hospital, facility, or physician. In these studies, confounding frequently occurs within or between clusters \cite{Neuhuasb, LiY, Salas, Bosco}. In our motivating example, we study the effect of erythropoietin-stimulating agent (ESA) administration on raising hemoglobin (Hgb) levels among patients receiving hemodialysis for end-stage kidney disease from multiple dialysis facilities. Facility indicators of quality of clinical care, such as the percentage of facility patients receiving dialysis via a central venous catheter, are likely between-cluster confounders, while patients' responsiveness to ESA (i.e., change in hemoglobin level) is likely a within-cluster confounder. When within- or between-cluster confounding is unmeasured or unadjusted for, most statistical methods are invalid and can give biased effect estimates. However, it is less known that some statistical methods are more robust to unmeasured confounding and give less biased effect estimates than others. In this study, we aim to quantify the bias of effect estimates obtained from the instrumental variable analysis (IVA) and several alternative methods in the presence of between- or within-cluster unmeasured confounding.

\begin{figure}
\centering
\includegraphics[width=12cm, height=6cm]{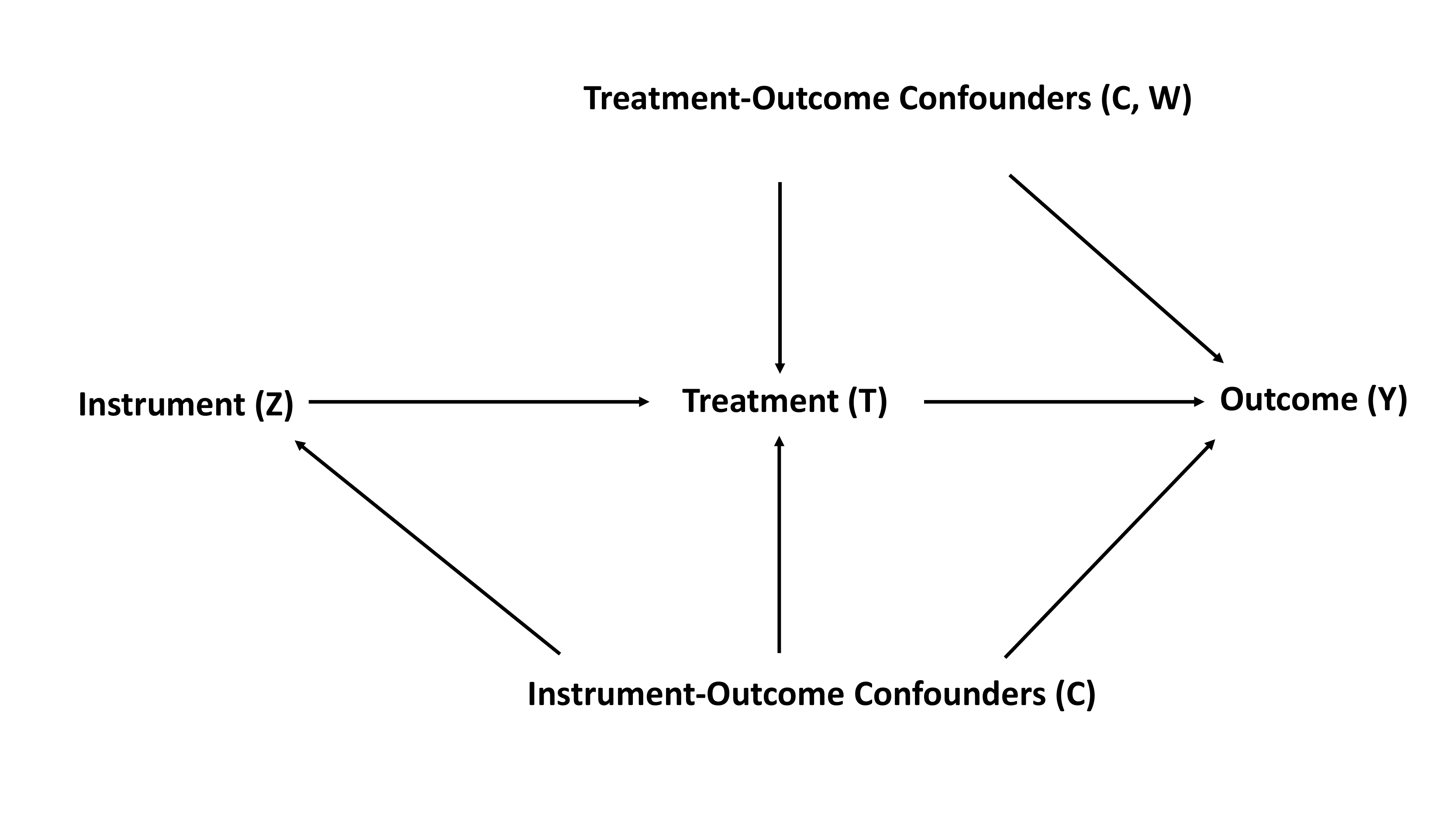}
\caption{Assumptions for Valid IV: (a) association between instrument and treatment; (b) no unmeasured IV-outcome confounders; (c) no direct effect of instrument on outcome. $Z$ for intrument, $T$ for treatment, $Y$ for outcome, $C$ for measured confounders, $W$ for unmeasured within-cluster confounders for the T-Y association.}
   \label{fig:1}
\end{figure}

IVAs are popular choices for obtaining effect estimates robust to unmeasured confounding. Treatment preferences of clusters (e.g., physician practices) are the most frequently used instruments in the literature (i.e., preference-based instruments) \cite{Korn, Chen, Davies, Brookhart}. Out of $187$ comparative effectiveness research studies that used IVA between year 1990 and 2011, about half of the instruments were preference-based instruments \cite{Garabedian}. Different practices prefer different treatment dose (percentages) even after sufficient adjustments for patient/practice heterogeneity. Part of the variation in treatment dose preferences may arise from differential group policies, insurance coverage, patient/physician knowledge/preference, and may be random such that it is independent of unmeasured confounders (e.g., patients' disease severity) after adjustment for measured confounders. The preference-based IVA aims to use the ``random'' component of  variation to obtain treatment effect estimates. For an IV to be valid, it must be associated with the exposure of interest, independent of unmeasured confounders and must have no direct effect on the outcome, conditional on measured covariates (Figure \ref{fig:1}). Previously, we derived a bias formula when assumptions for valid preference-based IVs are met \cite{LiY}. In this study, we generalize and derive bias formulae of preference-based IV estimators in the more common and realistic scenarios when IV assumptions may or may not be satisfied in the presence of unmeasured between-cluster and/or within-cluster confounders.

It is known that IVA is advantageous in handling unmeasured confounding; however, it is less known that other popular analytic methods may give less biased effect estimates than IVA depending on the nature of unmeasured confounding. For comparisons with IVA, we choose three commonly used models: ordinary least squares regression (OLS), linear mixed models (LMM) and fixed effect models (FE) to estimate the effect of a continuous exposure (e.g., treatment dose) on a continuous outcome. Previous work has focused on the importance of assumptions being satisfied in order for these statistical methods to generate valid effect estimates. However, as we demonstrate in this paper, when between- and within-cluster confounding are not fully controlled, most of these methods are likely to be invalid. And the robustness of these methods towards bias differs depending on the extent of unmeasured within-cluster confounding relative to unmeasured between-cluster confounding. Our prior work derived bias formula for IVA when unmeasured within-cluster confounding exists \cite{LiY}; in this study, we derive bias formulae for all four methods when between-cluster and/or within-cluster confounding exits. Our study of the bias patterns and factors that impact the magnitudes of the bias for these methods will: (a) provide evidence in selecting better methods to combat the dominant types of unmeasured confounders; (b) help appropriately interpret statistical results in the context of unmeasured confounding; and (c) assist in detecting the presence of unmeasured confounders.

As an illustrating example, we use these methods to estimate the effect of ESA on Hgb levels among patients receiving hemodialysis for end-stage kidney disease using data from the Dialysis Outcomes and Practice Patterns Study (DOPPS), an international prospective cohort study \cite{Pisoni}.

\section{Between- and Within-Cluster Confounders}
We aim to estimate the effect of an exposure ($T$) on outcome ($Y$) free of confounding. We categorize all unmeasured confounders of the $T-Y$ relationship into between-cluster confounders (denoted by $B$) and within-cluster confounders (denoted by $W$) \cite{Neuhuasb, LiY}. Between-cluster confounders $B_i$ are cluster-specific and identical for any patient $j$ in each cluster $i$; but $B_i$s likely differ across different clusters. Within-cluster confounders ($W_{ij}$) likely have different values for different subjects $j$'s within each cluster $i$; but their cluster-specific means are identical for all $i$. For example, when the cluster is the dialysis facility, between-cluster confounders can be for-profit status, or nurse/patient ratio of dialysis facilities. Many confounders tend to be both between-cluster and within-cluster confounders (e.g., patients' age). For these confounders, we decompose them into between-cluster components (e.g., facility mean age) and within-cluster components (e.g., a patient's age - facility mean age). The decomposition of unmeasured confounders helps us examine how each type affects the $T-Y$ relationship and the assumptions required for valid statistical methods. We let $C_{ij}$ denote all adjusted (measured) confounders. Here $W_{ij}$, $B_i$ and $C_{ij}$ are $K_w$, $K_b$ and $K_c$ dimensional vectors, respectively. We assume $W_{ij}$ and $B_i$ represent residual within-cluster and between-cluster components of the unadjusted confounders after controlling for $C_{ij}$ (i.e., $(W_{ij}, B_i) \perp C_{ij}$).

\section{Methods}
Our objective is to examine the effect of a continuous exposure (i.e., treatment dose) on a continuous outcome that is free of confounding. First, we describe the preference-based IVA method and three other methods to compare, their estimators and assumptions for valid effect estimates. We then assess possible violations of these assumptions when between-cluster or within-cluster unmeasured confounding exists.

\subsection{\underline{Preference-Based Instrumental Variable Approach}}
The models for the preference-based IVA are expressed below as two simultaneous equations:
\begin{eqnarray}
T_{ij} &=& \gamma_i + C_{ij}^{'*} \alpha_{Ic}^{*} + e_{ij}^t, \label{truetiv1} \\
Y_{ij} &=& \beta_{I} T_{ij} +  C_{ij}^{'} \beta_{Ic} + v_{i} + e_{ij}^y, \label{trueyiv1}
\end{eqnarray}
where $j=1, 2, \ldots, n_i$ for each $i$ and $i=1, 2, \ldots, m$, $\beta_{I}$ is the parameter of interest denoting the effect of $T$ on $Y$; $\gamma_i$ represents the random exposure level at cluster $i$; $\alpha_{Ic}^{*}= (\alpha_{2Ic}, \cdots, \alpha_{K_cIc})^{'}$ represents the effects of $C_{ij}^{*} = (C_{2ij}, \ldots, C_{K_cij})^{'}$ on $T_{ij}$; and $\beta_{Ic} = (\beta_{1Ic}, \ldots, \beta_{K_cIc})^{'}$ corresponds to the effects of $C_{ij} = (C_{1ij}, \ldots, C_{K_cij})^{'}$ on $Y_{ij}$. Note that $C_{1ij}=1$ denoting the intercept and $C_{2ij}, \ldots, C_{K_cij}$ for measured confounders. Also note that different from $C_{ij}$, $C_{ij}^{*}$ does not include intercept. In addition, different from LMM, $\gamma_i$ can be arbitrarily correlated with $C_{ij}^{'*}$. Without loss of generality, we assume cluster size $n_i=n$ for any $i$ for notational simplicity in the following.

We assume the between-cluster errors (or random effects) $v_i$ and within-cluster errors $e_{ij}^t$ and $e_{ij}^y$ are identically, independently distributed (i.i.d.) and follow normal distributions such that $v_{i} \sim N(0, \sigma_v^2)$, $e_{ij}^t \sim N(0, \sigma_{et}^2)$, and $e_{ij}^y \sim N(0, \sigma_{ey}^2)$, where $v_i \perp (e_{ij}^t, e_{ij}^y)$ . Here $v_i$ captures the intra-cluster correlation for the outcome.  Let $\xi_i = v_{i} J_n + e_{i}^y$ where $J_n$ is a $n \times 1$ vector of ones, and $e_{i}^y = (e_{i1}^y, \cdots, e_{in}^y)^{'}$. Then we have $\xi_i \sim N(0, \Omega)$ with $\Omega = \sigma_v^2 I_n + \sigma_{ey}^2 J_nJ_n^{'}$ where $I_n$ is an identity matrix with rank $n$. Other assumptions for valid IVA model equations include: $(v_{i}, e_{ij}^t, e_{ij}^y) \perp C_{ij}$ and $e_{ij}^y \perp T_{ij}$. Note that the assumption of $e_{ij}^t \perp e_{ij}^y$ is not required for valid IVA.

Let $\eta_I = (\beta_I, \beta_{1Ic}, \ldots, \beta_{K_cIc})^{'}$. With preference-based IVA, the two-stage generalized least squares estimator of $\eta_I$ is given by
\begin{eqnarray}
\widehat{\eta}_{I} = \left(\sum_{i=1}^m \widehat{O}_i^{'} \widehat{\Omega}^{-1} \widehat{O}_i \right)^{-1} \left(\sum_{i=1}^m \widehat{O}_i^{'} \widehat{\Omega}^{-1} Y_i \right), \label{2sgls}
\end{eqnarray}
where $\widehat{\Omega}$ is an estimate of $\Omega$ and $\widehat{O}_i=(\widehat{O}_{i1}, \ldots, \widehat{O}_{in})^{'}$ with $\widehat{O}_{ij} = (\widehat{T}_{ij}, C_{1ij}, \cdots, C_{K_cij})^{'}$. $\widehat{T}_{ij}$ is the predicted $T$ obtained from Equation ($\ref{truetiv1}$) using OLS estimation by regressing $T_{ij}$ on $C_{ij}^{'}$ and $Z_i^{'}$, where $Z_i$ is an $m \times 1$ indicator vector with its elements being I$({\ell=i}) = 1$ (if $\ell = i$) or $0$ (if $\ell \neq i$) where $\ell=1, \cdots, m$ representing any potential cluster memberships. The estimator for the coefficients of $Z_i$ is the least squares dummy variable estimator in economics \cite{Baltagi, Hsiao, Wooldridge}; it is also the same estimator of the coefficients in the fixed effect regression model in biostatistics and economics \cite{Allison, Baltagi, Hsiao, Wooldridge}. The estimator of $\beta_I$ is given by $\widehat{\beta}_{I} = (1, 0, \ldots, 0) \times \widehat{\eta}_{I}$. The variance of $\widehat{\eta}_{I}$ is estimated by: $\mbox{Var}(\widehat{\eta}_{I}) = \left(\sum_{i=1}^m \widehat{O}_i^{'} \widehat{\Omega}^{-1} \widehat{O}_i \right)^{-1}.$ Hence, $\mbox{Var}(\widehat{\beta}_{I}) = (1, 0, \ldots, 0)\mbox{Var}(\widehat{\eta}_{I})(1, 0, \ldots, 0)^{'}$. We obtain $\widehat{\Omega}$ using the procedure provided in our prior work \cite{LiY}.

The preference-based IVA aims at taking advantage of the random component of the treatment dose assignment to obtain valid effect estimates. Conditional on measured confounders, this random component is independent of unmeasured confounders, may arise from differential cluster policies or preference and represented by $\gamma_i$ Equation (\ref{truetiv1}). The IVA assumptions require that $\gamma_i$ does not have a direct effect on $Y$ and that $\gamma_i$ must be independent of the unmeasured confounders (conditional on measured covariates). This implies that $\gamma_i$ needs to be independent of $v_i, e_{ij}^t$ and $e_{ij}^y$. Additional IVA assumption includes that $\gamma$ is positively associated with the treatment dose received \cite{LiY}.

When unmeasured within-cluster confounders $W_{ij}$ are present, $W_{ij}$ are absorbed by $e_{ij}^t$ and $e_{ij}^y$ such that $e_{ij}^t = W_{ij}^{'} \alpha_w + \varepsilon_{ij}^t$ and $e_{ij}^y = W_{ij}^{'} \beta_w + \varepsilon_{ij}^y$ with $\alpha_w$ and $\beta_w$ representing the effects of $W$ on $T$ and $Y$, respectively. This results in Cov$(e_{ij}^t, e_{ij}^y) \neq 0$ and subsequently Cov$(T_{ij}, e_{ij}^y) \neq 0$, even if $\varepsilon_{ij}^t \bot \varepsilon_{ij}^y$. This non-zero correlation between $T_{ij}$ and $e_{ij}^y$ can result in biased estimates of $\beta_I$ when we only fit the single model equation (\ref{trueyiv1}). However, in IVA, when $W_{ij}$ are present, because of $\gamma_i \perp W_{ij}$, $\gamma_{i} \perp (e_{ij}^t, e_{ij}^y, v_i)$ remains true after adjusting for $C_{ij}$; and hence IVA assumptions are not violated. When unmeasured between-cluster confounders $B_i$ is present, $B_i$ is absorbed by $\gamma_i$ and $v_{i}$ such that $\gamma_i = B_i^{'}\alpha_b + r_{0i}$ and $v_i = B_i^{'}\beta_b + u_{0i}$ with $\alpha_b$ and $\beta_b$ representing the effect of $B$ on $T$ and $Y$, respectively. Subsequently, this leads to Cov$(\gamma_i, v_{i}) \neq 0$, even if $r_{0i} \bot u_{0i}$. This violates the $\gamma_i \perp v_{i}$ assumption required for a valid IVA and can lead to invalid IVA estimates. Here, $B$ is also termed as IV-outcome confounders in literature \cite{Garabedian}.

The causal interpretation of IVA estimators are described in further details by us and others \cite{LiY, Baiocchi0, Imbens, Sargan}. Briefly, to point identify the treatment effect, we need to make the additional assumptions of either monotonicity \textit{or} homogeneous effects. With the monotonicity assumption, we can interpret the IV estimate as the causal treatment effect for the compliers, the subgroup of patients who would adopt the treatment dose suggested by the instrument. With the homogeneous effect assumption, we can interpret the IVA estimate as the average causal effect for the whole population. In this manuscript, we assume homogeneous treatment dose effect for simulations, but make no such an assumption for the data analyses since the truth is unknown.

\subsection{Alternative Methods}
\subsubsection{\underline{Ordinary Least Squares Regression (OLS)} }
The model for OLS regression can be expressed as:
\begin{eqnarray}
Y_{ij} = \beta_{O} T_{ij} +  C_{ij}^{'} \beta_{Oc} + \varsigma_{ij}^y, \label{ols}
\end{eqnarray}
where $\beta_{O}$ is the parameter of interest representing the effect of $T$ on $Y$, $\varsigma_{ij}^y$ is the random error, and $\beta_{Oc} = (\beta_{1Oc}, \ldots, \beta_{K_cOc})^{'}$ for the effects of $C_{ij}$ on $Y_{ij}$. Let $\eta_O = (\beta_O, \beta_{1Oc}, \ldots, \beta_{K_cOc})^{'}$, $O_{ij} = (T_{ij}, C_{1ij}, \cdots, C_{K_cij})^{'}$, $Y_i = ({Y_{i1}, Y_{i2}, \ldots, Y_{in}})^{'}$ and $T_i = ({T_{i1}, T_{i2}, \ldots, T_{in}})^{'}$. We assume that $\varsigma_{ij}^y \sim N(0, \sigma_{\varsigma}^2)$. The OLS estimator ($\hat{\eta}_{O}$) of $\eta_{O}$ is:
\begin{eqnarray}
\widehat{\eta}_{O} = \left(\sum_{i=1}^m O_i^{'} O_i \right)^{-1} \left(\sum_{i=1}^m O_i^{'} Y_i \right) . \label{olsest}
\end{eqnarray}
The OLS estimator of $\beta_O$ is then given by $\widehat{\beta}_{O} = (1, 0, \ldots, 0) \times \widehat{\eta}_{O}$. The variance of $\widehat{\eta}_{O}$ is estimated by: ${Var}(\widehat{\eta}_{O}) = \left(\sum_{i=1}^m O_i^{'} O_i \right)^{-1}\sigma_{\varsigma}^2$. Hence, $\mbox{Var}(\widehat{\beta}_{O}) = (1, 0, \ldots, 0)\mbox{Var}(\widehat{\eta}_{O})(1, 0, \ldots, 0)^{'}$.

The assumptions required for the OLS regression to obtain a valid estimate of $\beta_{O}$ are: $\varsigma_{ij}^y$s are i.i.d. and $ \varsigma_{ij}^y \perp (T_{ij}, C_{kij})$ for any $k$. In clustered data, $\varsigma_{ij}^y$s are correlated, and hence the i.i.d. assumption is violated. Furthermore, when unmeasured between-cluster or within-cluster confounders exist, the assumption $\varsigma_{ij}^y \perp T_{ij}$ is no longer valid.

\subsubsection{\underline{Fixed Effect Regression Model (FE)}}
The fixed-effect regression model is expressed as:
\begin{eqnarray}
Y_{ij} = \beta_F T_{ij} + C_{ij}^{'}\beta_{Fc} + \mu_2 + \ldots + \mu_m + \varepsilon_{ij}^y,  \label{fe}
\end{eqnarray}
where $\beta_F$ is the parameter of interest for the effect of $T$ on $Y$; $\mu_2, \ldots, \mu_m$ are fixed effects, representing cluster-specific effects for cluster 2 through $m$ and $\varepsilon_{ij}^y$ is the random error with $\varepsilon_{ij}^y \sim N(0, \sigma_\varepsilon^2)$. Let $\overline{Y}_i = \frac{1}{n_i} \sum_{j=1}^{n_i} Y_{ij}$, $\overline{T}_i = \frac{1}{n_i} \sum_{j=1}^{n_i} T_{ij}$, $\overline{C}_{ki} = \frac{1}{n_i} \sum_{j=1}^{n_i} C_{kij}$, and $\overline{\varepsilon}_i = \frac{1}{n_i}\sum_{j=1}^{n_i} \varepsilon_{ij}$. To obtain the fixed effect estimator, we first subtract $\overline{Y}_i$ from both sides of the model (\ref{fe}) and obtain the following model:
\begin{eqnarray}
Y_{ij}^{\dag} = \beta_F T_{ij}^{\dag} + C_{ij}^{'{\dag}}\beta_{Fc} + \varepsilon_{ij}^{y\dag},  \nonumber
\end{eqnarray}
where $Y_{ij}^{\dag} = Y_{ij} - \overline{Y}_i$, $T_{ij}^{\dag} = T_{ij} - \overline{T}_i$, $C_{kij}^{\dag} = C_{kij} - \overline{C}_{ki}$ and $\varepsilon_{ij}^{y\dag} = \varepsilon_{ij}^y  - \overline{\varepsilon}_i$. Note that here $\mu_2, \ldots, \mu_m$, the intercept $C_{1ij}$, and any measured between-cluster confounders are eliminated by this transformation. Hence, $C_{ij}^{\dag}$ excludes the intercept and measured between-cluster confounders. Let $O_{ij}^{\dag} = (T_{ij}^{\dag}, \ldots, C_{kij}^{\dag}, \ldots)$ and $\hat{\delta}_F = (\beta_F, \beta_{1Fc}, \ldots, \beta_{K_cFc})$. The fixed effect estimator is given by
\begin{eqnarray}
\hat{\delta}_F = \left(\sum_{i=1}^m O_i^{'\dag} O_i^{'\dag} \right)^{-1} \left(\sum_{i=1}^m O_i^{'\dag} Y_i^{'\dag} \right). \nonumber
\end{eqnarray}
The fixed-effect estimator is also named as within-cluster estimator and is numerically identical to least squares dummy variable estimator. The fixed-effect estimator of $\beta_F$ is given by $\hat{\beta}_F = (1, 0, \ldots, 0) \times \hat{\delta}_F$. The variance of $\hat{\beta}_F$ is estimated by $Var(\hat{\beta}_F) = \left(\sum_{i=1}^m O_i^{'\dag} O_i^{'\dag} \right)^{-1} \sigma_{\varepsilon}^2$. Hence, $Var(\hat{\beta}_F) = (1, 0, \ldots, 0) Var(\hat{\delta}_F)(1,0, \ldots, 0)^{'}$.

The assumptions required for the fixed-effect regression to obtain a valid estimate of $\beta_F$ are: $\varepsilon_{ij}^{y}$ are i.i.d. and  $\varepsilon_{ij}^{y} \perp (T_{ij}, C_{kij})$ for any $k$. Since the statistical inference for the fixed-effect regression model is built upon conditioning being in the same cluster, subjects in the same cluster are still i.i.d. even if there is a correlation among subjects who belong to the same cluster. Subsequently, unmeasured between-cluster confounders $B$ do not have any impact on the model assumptions and are absorbed by the fixed effects $\mu_2, \ldots, \mu_m$. However, when unmeasured within-cluster confounders $W$ exist, the assumption $\varepsilon_{ij}^{y} \perp T_{ij}$ no longer holds.

\subsubsection{\underline{Linear Mixed Model (LMM)}}
The LMM is written as follows:
\begin{eqnarray}
Y_{ij} = \beta_{L} T_{ij} +  C_{ij}^{'} \beta_{Lc} + d_{0i} + \chi_{ij}^y, \label{lm}
\end{eqnarray}
where $\beta_{L}$ and $\beta_{Lc} = (\beta_{1Lc}, \ldots, \beta_{K_cLc})^{'}$ correspond to the effects of $T_{ij}$ and $C_{ij}^{'}$ respectively. Note that $d_{0i}$ refers to the random effect or between-cluster error, and $\chi_{i}^y = (\chi_{i1}^y, \cdots, \chi_{in}^y)^{'}$ refers to the within-cluster error. $\chi_{ij}^y$ and $d_{0i}$ represent the totality of within-cluster covariates and between-cluster covariates, respectively, omitted from the model that are orthogonal to covariates already in the model \cite{Neuhaus2}. The random effect $d_{0i}$ accommodates the intra-cluster correlation. We assume that $d_{0i}$ and $\chi_{ij}^y$ are i.i.d., and $d_{0i} \sim N(0, \sigma_d^2)$ and $\chi_{ij}^y \sim N(0, \sigma_{\chi}^2)$. Let $\zeta_i = d_{0i} J_n + \chi_{i}^y$. We have $\zeta_i \sim N(0, \Phi)$ with $\Phi = \sigma_d^2 I_n + \sigma_{\chi}^2 J_nJ_n^{'}$.

Let $\eta_L = (\beta_L, \beta_{1Lc}, \ldots, \beta_{K_cLc})^{'}$. The maximum likelihood estimator of $\eta_{L}$ is given by
\begin{eqnarray}
\widehat{\eta}_{L} = \left(\sum_{i=1}^m O_i^{'} \widehat{\Phi}^{-1} O_i \right)^{-1} \left(\sum_{i=1}^m O_i^{'} \widehat{\Phi}^{-1} Y_i \right), \label{lmest}
\end{eqnarray}
where $\widehat{\Phi}$ is an estimate of $\Phi$. The estimator of $\beta_L$ is given by $\hat{\eta}_{L} = (1, 0, \ldots, 0) \times \hat{\eta}_{L}$. The variance estimate is $Var(\widehat{\eta}_L) = \left(\sum_{i=1}^m O_i^{'} \widehat{\Phi}^{-1} O_i \right)^{-1}$. Hence, $\mbox{Var}(\widehat{\beta}_{L}) = (1, 0, \ldots, 0)\mbox{Var}(\widehat{\eta}_{L})(1, 0, \ldots, 0)^{'}$.

The assumptions required for the LMM to obtain unbiased estimates of $\beta_L$ include $(T_{ij}, C_{kij}) \perp \chi_{ij}^y$ (namely, the level-1 independence \cite{Ebbes}), $(T_{ij}, C_{kij}) \perp d_{0i}$ (namely, the level-2 independence \cite{Ebbes}) for any $k$, and $d_{0i} \perp \chi_{ij}^y$.  When unmeasured within-cluster confounding exists, it is absorbed by $\chi_{ij}^y$.  This induces correlation between $T_{ij}$ and $\chi_{ij}^y$ and hence violates the $T_{ij} \perp \chi_{ij}^y$ assumption. When unmeasured between-cluster confounding exits, it is absorbed by $d_{0i}$. This induces correlation between $T_{ij}$ and $d_{0i}$ and hence violates the $T_{ij} \perp d_{0i}$ assumption.  When both unmeasured between-cluster and within-cluster confounders exist, both level-1 and level-2 independence assumptions for valid LMM are no longer valid.

\section{Bias in the Presence of Unmeasured Confounders}
In this section we first specify true models for $T$ and $Y$ and then derive the expression of asymptotic bias of the four effect estimators described above when the exposure and outcome are continuous and unmeasured within-cluster and/or between-clustering confounders exist. Note that the asymptotic bias is derived assuming the number of independent units (i.e., clusters) goes to infinity ($m\rightarrow\infty$). However, we will examine how well our asymptotic bias formulae approximate bias in finite samples when the number of clusters is finite through simulations in the next section.

\subsection{True Models}
LMMs are commonly used in clustered data settings to estimate the effect of a continuous exposure (i.e., treatment dose) on a continuous outcome. We assume that $T$ and $Y$ are generated by LMMs with measured confounders $C_{ij}$ and unmeasured confounders $W_{ij}$ and $B_i$ as follows:
\begin{eqnarray}
T_{ij} &=& a_{0i} + C_{ij}^{'}\alpha_{c} + W_{ij}^{'}\alpha_{w} + B_{i}^{'}\alpha_{b} + \epsilon_{ij}^t, \label{truet1} \\
Y_{ij} &=& b_{0i} + \beta T_{ij} +  C_{ij}^{'} \beta_{c} + W_{ij}^{'}\beta_{w} + B_{i}^{'}\beta_{b} + \epsilon_{ij}^y, \label{truey1}
\end{eqnarray}
where $\beta$ is the parameter of interest for the effect of $T$ on $Y$; $\alpha_{c}$, $\alpha_{w}$, $\alpha_{b}$, $\beta$, $\beta_c$, $\beta_w$ and $\beta_b$ are fixed effects; $a_{0i}$ and $b_{0i}$ are between-cluster random errors (or random effects); $\epsilon_{ij}^t$ and $\epsilon_{ij}^y$ are within-cluster random errors. We assume $a_{0i}$, $b_{0i}$, $\epsilon_{ij}^t$ and $\epsilon_{ij}^y$ are i.i.d., and $a_{0i} \sim N(0,\sigma_{a}^2)$, $b_{0i} \sim N(0, \sigma_{b}^2)$, $\epsilon_{ij}^t \sim N(0,\sigma_{\epsilon t}^2)$ and $\epsilon_{ij}^y \sim N(0,\sigma_{\epsilon y}^2)$. We also make the standard assumptions required in valid LMMs: $a_{0i} \perp (C_{kij}, \epsilon_{ij}^t, \epsilon_{ij}^y, W_{kij}, B_i)$, $b_{0i} \perp (a_{0i}, C_{kij}, T_{ij}, \epsilon_{ij}^t, \epsilon_{ij}^y, W_{kij}, B_i)$, $\epsilon_{ij}^t \perp (C_{kij}, W_{kij}, B_i)$ and $\epsilon_{ij}^y \perp (\epsilon_{ij}^t, C_{kij}, T_{ij}, W_{kij}, B_i)$ for any $k$.

Let $T_i = ({T_{i1}, T_{i2}, \ldots, T_{in}})^{'}$, $Y_i = ({Y_{i1}, Y_{i2}, \ldots, Y_{in}})^{'}$, $C_i = ({C_{i1}, C_{i2}, \ldots, C_{in}})^{'}$, $W_i = ({W_{i1}, W_{i2}, \ldots, W_{in}})^{'}$, $\epsilon_i^t = ({\epsilon_{i1}^t, \epsilon_{i2}^t, \ldots, \epsilon_{in}^t})^{'}$, and $\epsilon_i^y = ({\epsilon_{i1}^y, \epsilon_{i2}^y, \ldots, \epsilon_{in}^y})^{'}$. The true models can be expressed in a matrix form as follows:
\begin{eqnarray}
T &=& A_0 + C\alpha_{c} + W\alpha_{w} + B\alpha_{b}\otimes J_n + \epsilon^t , \label{truet2} \\
Y &=& B_0 + \beta T +  C\beta_{c} + W\beta_{w} + B\beta_{b}\otimes J_n + \epsilon^y, \label{truey2}
\end{eqnarray}
where $Y$, $T$, $C$, $W$, $\epsilon^t$, $\epsilon^y$, $A_0$ and $B_0$ consist of stacked elements of $Y_i$, $T_i$, $C_i$, $W_i$, $\epsilon_i^t$, $\epsilon_i^y$, $a_{0i}J_{n}$ and $b_{0i}J_{n}$ respectively. Note that $Y$, $T$, $\epsilon^t$, $\epsilon^y$, $A_0$ and $B_0$ are $mn \times 1$ vectors, $W$ and $C$ are $mn \times K_w$, and $mn \times K_c$ matrices respectively, and $B = (B_1, B_2, \ldots, B_m)^{'}$ is an $m\times K_b$ matrix.

\subsection{Asymptotic Bias}
With the true models described above, we derive and compare close-form expressions of asymptotic bias of the effect estimates obtained from the four methods when their model assumptions may or may not hold in the presence of unmeasured between-cluster or within-cluster confounders. Table \ref{Tab:1} summarizes the asymptotic bias formulae for these four methods under two scenarios: (1) when the number of clusters $m\rightarrow\infty$ and the cluster size is fixed at $n$; (2) when both $m\rightarrow\infty$ and the cluster size $n\rightarrow\infty$.

\begin{sidewaystable}
\begin{center}
\begin{tabular}{lcccccc }
\hline
Unmeasured & \multicolumn{4}{c}{Asymptotic Bias Formulae} \\
\hline
Confounder  & IV  & OLS & FE & LMM \\
\hline
 & \multicolumn{4}{c}{$m\rightarrow\infty$} \\
\hline
W &
$\frac{\alpha _{w}^{^{\prime }}V_{w}\beta _{w}/n}{%
\sigma _{a}^{2}+(\alpha _{w}^{^{\prime }}V_{w}\alpha _{w}+\sigma _{\epsilon
t}^{2})/n}$ &
$\frac{\alpha _{w}^{^{\prime }}V_{w}\beta _{w}}{%
\sigma _{a}^{2}+\alpha _{w}^{^{\prime }}V_{w}\alpha _{w}+\sigma _{\epsilon
t}^{2}}$  & $\frac{\alpha _{w}^{^{\prime }}V_{w}\beta _{w}}{%
\alpha _{w}^{^{\prime }}V_{w}\alpha _{w}+\sigma _{\epsilon t}^{2}}$ & $\frac{\alpha_w^{'}V_w\beta_w }{\sigma_a^2\frac{\sigma_{\chi e}^2}{\sigma_{\chi e}^2 + (n-1)\sigma_{de}^2} + (\alpha_w^{'}V_w\alpha_w + \sigma_{\epsilon t}^2)}$ \\
B & $\frac{\alpha _{b}^{^{\prime }}V_{b}\beta _{b}}{%
(\sigma _{a}^{2}+\alpha _{b}^{^{\prime }}V_{b}\alpha _{b})+\sigma _{\epsilon
t}^{2}/n}$ & $ \frac{\alpha _{b}^{^{\prime }}V_{b}\beta _{b}}{%
\sigma _{a}^{2}+\alpha _{b}^{^{\prime }}V_{b}\alpha _{b}+\sigma _{\epsilon
t}^{2}}$ & 0 & $\frac{\alpha_b^{'}V_b\beta_b}{(\sigma_a^2+\alpha_b^{'}V_b\alpha_b) + \sigma_{\epsilon t}^2\frac{\sigma_{\chi e}^2 + (n-1)\sigma_{de}^2}{\sigma_{\chi e}^2}}$ \\
W and B &
$\frac{\alpha _{b}^{^{\prime }}V_{b}\beta
_{b}+\alpha _{w}^{^{\prime }}V_{w}\beta _{w}/n}{(\sigma _{a}^{2}+\alpha
_{b}^{^{\prime }}V_{b}\alpha _{b})+(\alpha _{w}^{^{\prime }}V_{w}\alpha
_{w}+\sigma _{\epsilon t}^{2})/n}$ & $ \frac{\alpha _{b}^{^{\prime }}V_{b}\beta
_{b}+\alpha _{w}^{^{\prime }}V_{w}\beta _{w}}{\sigma _{a}^{2}+\alpha
_{b}^{^{\prime }}V_{b}\alpha _{b}+\alpha _{w}^{^{\prime }}V_{w}\alpha
_{w}+\sigma _{\epsilon t}^{2}}$ & $\frac{\alpha _{w}^{^{\prime }}V_{w}\beta _{w}}{%
\alpha _{w}^{^{\prime }}V_{w}\alpha _{w}+\sigma _{\epsilon t}^{2}}$ & $\frac{\alpha_b^{'}V_b\beta_b \frac{\sigma_{\chi e}^2}{\sigma_{\chi e}^2 + (n-1)\sigma_{de}^2} + \alpha_w^{'}V_w\beta_w }{(\sigma_a^2+\alpha_b^{'}V_b\alpha_b)\frac{\sigma_{\chi e}^2}{\sigma_{\chi e}^2 + (n-1)\sigma_{de}^2} + (\alpha_w^{'}V_w\alpha_w + \sigma_{\epsilon t}^2)}$ \\
\hline
& \multicolumn{4}{c}{$m\rightarrow\infty$ and $n\rightarrow\infty$ } \\
\hline
W &
0 &
$\frac{\alpha _{w}^{^{\prime }}V_{w}\beta _{w}}{%
\sigma _{a}^{2}+\alpha _{w}^{^{\prime }}V_{w}\alpha _{w}+\sigma _{\epsilon
t}^{2}}$  & $\frac{\alpha _{w}^{^{\prime }}V_{w}\beta _{w}}{%
\alpha _{w}^{^{\prime }}V_{w}\alpha _{w}+\sigma _{\epsilon t}^{2}}$ & $\frac{\alpha_w^{'}V_w\beta_w }{\alpha_w^{'}V_w\alpha_w + \sigma_{\epsilon t}^2}$ \\
B & $\frac{\alpha _{b}^{^{\prime }}V_{b}\beta _{b}}{%
\sigma _{a}^{2}+\alpha _{b}^{^{\prime }}V_{b}\alpha _{b}}$ & $ \frac{\alpha _{b}^{^{\prime }}V_{b}\beta _{b}}{%
\sigma _{a}^{2}+\alpha _{b}^{^{\prime }}V_{b}\alpha _{b}+\sigma _{\epsilon
t}^{2}}$ & 0 & 0 \\
W and B &
$\frac{\alpha _{b}^{^{\prime }}V_{b}\beta
_{b}}{\sigma _{a}^{2}+\alpha
_{b}^{^{\prime }}V_{b}\alpha _{b}}$ & $ \frac{\alpha _{b}^{^{\prime }}V_{b}\beta
_{b}+\alpha _{w}^{^{\prime }}V_{w}\beta _{w}}{\sigma _{a}^{2}+\alpha
_{b}^{^{\prime }}V_{b}\alpha _{b}+\alpha _{w}^{^{\prime }}V_{w}\alpha
_{w}+\sigma _{\epsilon t}^{2}}$ & $\frac{\alpha _{w}^{^{\prime }}V_{w}\beta _{w}}{%
\alpha _{w}^{^{\prime }}V_{w}\alpha _{w}+\sigma _{\epsilon t}^{2}}$ & $\frac{\alpha_w^{'}V_w\beta_w }{\alpha_w^{'}V_w\alpha_w + \sigma_{\epsilon t}^2}$ \\
\hline
\end{tabular}
\end{center}
\caption{Asymptotic Bias Formulae for Preference-based Instrumental Variable Analysis (IV), Ordinary Least Squares (OLS), Fixed Effect Models (FE) and Linear Mixed Models (LMM) in the Presence of Unmeasured Within- and/or Between-Cluster Confounders. $m$ and $n$ refer to the number of clusters and the cluster size. The true model is $T_{ij} = a_{0i} + C_{ij}^{'}\alpha_{c} + W_{ij}^{'}\alpha_{w} + B_{i}^{'}\alpha_{b} + \epsilon_{ij}^t$ and $Y_{ij} = b_{0i} + \beta T_{ij} +  C_{ij}^{'} \beta_{c} + W_{ij}^{'}\beta_{w} + B_{i}^{'}\beta_{b} + \epsilon_{ij}^y$. Note that $T$ for exposure, $Y$ for outcome, $C$ for measured confounders, W and B for unmeasured within- and between-cluster confounders respectively, $\sigma _{a}^{2}$, $\sigma _{b}^{2}$, $\sigma _{\epsilon t}^{2}$, $\sigma _{\epsilon y}^{2}$, $V_{b}$, $V_{w}$
denote the variances of $a_{0i}$, $b_{0i}$, $\epsilon _{ij}^{t}$, $\epsilon _{ij}^{y}$, $B_{i}$, and $W_{ij}$
for any subject $j$ in cluster $i$, $\sigma_{\chi e}^2$ and $\sigma_{de}^2$ refer to asymptotic variances of the random intercept and within-cluster error respectively for linear mixed models. }
  \label{Tab:1}
\end{sidewaystable}

To obtain these bias formulae, we make two assumptions to simplify the derivation process without lessening the generality of the formulae. First, we assume that the means of $W$, $B$, $Y$ and $T$ are all zeros because the means can only influence the estimates of the intercept but not the estimates for the parameter of interest, $\beta$. Second, we assume there are no measured confounders $C$ (including the intercept). To examine the impact of $C$ on the bias derivation, we first transform $Y$ and $T$ by pre-multiplying $M_C = I_{mn} - C(C^{'}C)^{-1}C^{'}$ to Equations ($\ref{truet2}$) and ($\ref{truey2}$) and obtain:
\begin{eqnarray}
T^{\ddag} &=& A_0 + W\alpha_{w} + B\alpha_{b}\otimes J_n + \epsilon^t, \nonumber  \\
Y^{\ddag} &=& B_0 + \beta T^{\ddag} + W\beta_{w} + B\beta_{b}\otimes J_n + \epsilon^y, \nonumber
\end{eqnarray}
where $T^{\ddag} = M_cT$ and $Y^{\ddag} = M_cY$ are projection errors of $Y$ and $T$ on the space spanned by $C$. Note that $M_c\epsilon^y=\epsilon^y$, $M_c\epsilon^t=\epsilon^t$, $M_cW=W$, and $M_cB=B$ because $C_{ij} \bot (\epsilon^t_{ij}, \epsilon^y_{ij}, W_{ij}, B_i, a_{0i}, b_{0i})$. The derived bias formulae only consist of the second moments of $W$, $B$, $\epsilon^t$ and $\epsilon^y$ and will not change by the process of projection. Hence, the transformation will not change any element in the bias formulae; and the bias formulae should be the same with or without adjusting $C$. This technique was successfully implemented in our prior paper \cite{LiY}. Through simulations, we will further confirm that the assumptions of mean zeros and no $C$ will not influence the bias formulae.

\subsubsection{\underline{Preference-Based Instrumental Variable Approach}}
In the absence of $C$, the two-stage generalized least squares estimator of $\beta$ in (\ref{2sgls}) can then be simplified to
\begin{eqnarray}
\widehat{\beta}_{I} &=& \left(\sum_{i=1}^m \widehat{T}_i^{'} \widehat{\Omega}^{-1} \widehat{T}_i \right)^{-1} \left(\sum_{i=1}^m \widehat{T}_i^{'} \widehat{\Omega}^{-1} Y_i \right), \label{IVE}
\end{eqnarray}
where ${\widehat{T}_i} = J_nJ_n^{'}T_i/n$ and $\widehat{\Omega }^{-1}$ is an estimator for $\Omega ^{-1}$. When both unmeasured between- and within-cluster
confounders exist, $\Omega =(\beta _{b}^{^{\prime }}V_{b}\beta _{b}+\sigma _{b}^{2})I_{n}+(\beta
_{w}^{^{\prime }}V_{w}\beta _{w}+\sigma _{\epsilon
y}^{2})J_{n}J_{n}^{^{\prime }}$ and $\Omega ^{-1}=H_I I_{n}- H_I H_J J_{n}J_{n}^{^{\prime }}$,
where $H_I = \frac{1}{\beta _{w}^{^{\prime }}V_{w}\beta _{w}+\sigma
_{\epsilon y}^{2}}$, $H_J = \frac{\beta _{b}^{^{\prime }}V_{b}\beta
_{b}+\sigma _{b}^{2}}{(\beta _{w}^{^{\prime }}V_{w}\beta _{w}+\sigma
_{\epsilon y}^{2})+n(\beta _{b}^{^{\prime }}V_{b}\beta _{b}+\sigma _{b}^{2})}$, with $\sigma _{b}^{2}$, $\sigma _{\epsilon y}^{2}$, $V_{b}$, and $V_{w}$
representing the variance of $b_{0i}$, $\epsilon _{ij}^{y}$, $B_{i}$, and $W_{ij}$
for any $i,j$, respectively.

When unmeasured between- and within-cluster confounders exist, as $m\rightarrow\infty$, as proved in the Appendix, the bias of $\widehat{\beta}_{I}$ can be approximated as
\begin{eqnarray}
\widehat{\beta }_{I}-\beta   \rightarrow_p \frac{\alpha _{b}^{^{\prime }}V_{b}\beta
_{b}+\alpha _{w}^{^{\prime }}V_{w}\beta _{w}/n}{(\sigma _{a}^{2}+\alpha
_{b}^{^{\prime }}V_{b}\alpha _{b})+(\alpha _{w}^{^{\prime }}V_{w}\alpha
_{w}+\sigma _{\epsilon t}^{2})/n}. \nonumber
\end{eqnarray}
When both $m\rightarrow\infty$ and $n\rightarrow\infty$, $\widehat{\beta}_{I} - \beta  \rightarrow_p \frac{\alpha_b^{'}V_b\beta_b }{\sigma_a^2+\alpha_b^{'}V_b\alpha_b}$, a function of unmeasured between-cluster confounders but not within-cluster confounders.

When unmeasured within-cluster confounders but not between-cluster confounders exist, as  $m\rightarrow\infty$, the bias of $\widehat{\beta}_{I}$ can be simplified as
\begin{eqnarray}
\widehat{\beta }_{I}-\beta  \rightarrow_p \frac{\alpha _{w}^{^{\prime }}V_{w}\beta _{w}/n}{%
\sigma _{a}^{2}+(\alpha _{w}^{^{\prime }}V_{w}\alpha _{w}+\sigma _{\epsilon
t}^{2})/n}, \nonumber
\end{eqnarray}
where $\sigma _{a}^{2}$ and $\sigma _{\epsilon
t}^{2}$ denote the variance of $a_{0i}$ and $\epsilon _{ij}^{t}$ for any $%
i,j $, respectively.
This bias formula was first derived in our prior paper \cite{LiY}. As stated previously, when only unmeasured within-cluster confounders exist, the assumptions for the IVA are not violated and the IVA is valid. Nonetheless, finite sample bias still exits \cite{LiY}. However, when both $m\rightarrow\infty$ and $n\rightarrow\infty$, $\widehat{\beta}_{I} - \beta  \rightarrow_p 0$. Hence, unmeasured within-cluster confounders lead to finite-sample bias but not asymptotic bias. When the number of clusters and cluster sizes are large, the bias of instrumental variable estimates becomes negligible.

When unmeasured between-cluster but not within-cluster confounding exist, as $m\rightarrow\infty$, the bias of $\widehat{\beta}_{I}$ can be simplified as
\begin{eqnarray}
\widehat{\beta }_{I}-\beta   \rightarrow_p \frac{\alpha _{b}^{^{\prime }}V_{b}\beta _{b}}{%
(\sigma _{a}^{2}+\alpha _{b}^{^{\prime }}V_{b}\alpha _{b})+\sigma _{\epsilon
t}^{2}/n}. \nonumber
\end{eqnarray}
When both $m\rightarrow\infty$ and $n\rightarrow\infty$, $\widehat{\beta }_{I}-\beta  \rightarrow_p \frac{\alpha
_{b}^{^{\prime }}V_{b}\beta _{b}}{\sigma _{a}^{2}+\alpha _{b}^{^{\prime
}}V_{b}\alpha _{b}}$, which is the same as the asymptotic bias when both unmeasured between- and within-cluster confounders exist. Hence, unmeasured between-cluster confounders result in both finite and asymptotic bias.

\subsubsection{\underline{Ordinary Least Squares Regression}}
In the absence of $C$, the OLS estimator in ($\ref{olsest}$) can be simplified to:
\begin{eqnarray}
\widehat{\beta }_{O}=\left( \sum_{i=1}^{m}T_{i}^{\prime }T_{i}\right)
^{-1}\left( \sum_{i=1}^{m}T_{i}^{\prime }Y_{i}\right). \nonumber
\end{eqnarray}
When unmeasured between- and within-cluster confounders exist, as shown in The Appendix, with $m\rightarrow\infty$, the asymptotic bias of $\widehat{\beta }_{O}$ is given as
\begin{equation*}
\widehat{\beta }_{O}-\beta  \rightarrow_p \frac{\alpha _{b}^{^{\prime }}V_{b}\beta
_{b}+\alpha _{w}^{^{\prime }}V_{w}\beta _{w}}{\sigma _{a}^{2}+\alpha
_{b}^{^{\prime }}V_{b}\alpha _{b}+\alpha _{w}^{^{\prime }}V_{w}\alpha
_{w}+\sigma _{\epsilon t}^{2}}.
\end{equation*}%
Note that this bias formula does not depend on the cluster size $n$. Both unmeasured between- and within-cluster confounders contribute to the bias.

When unmeasured within-cluster confounders but not between-cluster confounders exist, as $m\rightarrow\infty$, the bias formula is simplified to
\begin{equation*}
\widehat{\beta }_{O}-\beta   \rightarrow_p \frac{\alpha _{w}^{^{\prime }}V_{w}\beta _{w}}{%
\sigma _{a}^{2}+\alpha _{w}^{^{\prime }}V_{w}\alpha _{w}+\sigma _{\epsilon
t}^{2}}.
\end{equation*}%

When unmeasured between-cluster but not within-cluster confounding exist, as $m\rightarrow\infty$, the bias formula is simplified to
\begin{equation*}
\widehat{\beta }_{O}-\beta  \rightarrow_p \frac{\alpha _{b}^{^{\prime }}V_{b}\beta _{b}}{%
\sigma _{a}^{2}+\alpha _{b}^{^{\prime }}V_{b}\alpha _{b}+\sigma _{\epsilon
t}^{2}}.
\end{equation*}%
Note that, when only unmeasured between-cluster confounders exist, the absolute value of the asymptotic bias of $\widehat{%
\beta }_{O}$ is smaller than that of $\widehat{\beta }_{I}$ for any $n\geq 2$
and the difference between asymptotic bias of $\widehat{\beta}_{O}$ and $\widehat{
\beta }_{I}$ becomes larger as the cluster size $n$ grows.

\subsubsection{\underline{Fixed Effect Model}}
In the absence of $C$, the FE estimator in (\ref{fe}) can be simplified to
\begin{eqnarray}
\hat{\beta}_{F} = \left(\sum_{i=1}^m T_i^{'} (I_n - Q_n) T_i \right)^{-1} \left(\sum_{i=1}^M T_i^{'} (I_n - Q_n) Y_i \right). \nonumber
\end{eqnarray}
where $Q_n = J_n(J_n^{'}J_n)^{-1}J_n^{'} = J_nJ_n^{'}/n$ as the orthogonal projection matrix of $J_n$.

When unmeasured between- and within-cluster confounders exist, as shown in the Appendix, as $m\rightarrow\infty$, the bias of $\hat{\beta}_{F}$ can be approximated as
\begin{equation*}
\widehat{\beta }_{F}-\beta  \rightarrow_p \frac{\alpha _{w}^{^{\prime }}V_{w}\beta _{w}}{%
\alpha _{w}^{^{\prime }}V_{w}\alpha _{w}+\sigma _{\epsilon t}^{2}}.
\end{equation*}%
Like the OLS case, this expression does not
depend on the size of $n$ and only depends on unmeasured within-cluster confounders.

When unmeasured within-cluster but not between-cluster confounders exist, the bias formula remains unchanged.

When unmeasured between-cluster but not within-cluster confounders exist, $\widehat{\beta }_{F}-\beta  \rightarrow_p 0$.

This implies that only unmeasured within-cluster confounders will result in asymptotic bias in the FE estimator (i.e., as $m\rightarrow\infty$). It is quite intuitive since $\mu _{i}$
and the intercept term in Equation (\ref{fe}) completely control for unobserved between-cluster confounders and exhaust the degree of freedom at the
cluster level.

\subsubsection{\underline{Linear Mixed Model}}
In the absence of $C$, the LMM estimator in Equation ($\ref{lmest}$) can be simplified as
\begin{eqnarray}
\hat{\beta}_{L} = \left(\sum_{i=1}^M T_i^{'} \widehat{\Phi}^{-1} T_i \right)^{-1} \left(\sum_{i=1}^M T_i^{'} \widehat{\Phi}^{-1} Y_i \right), \nonumber
\end{eqnarray}
where  $\Phi $ is the same as $\Omega $ in Equation ($\ref{IVE}$). When unmeasured between- and within-cluster confounders exist, the cluster-level error term in ($\ref{lm}$) $d_{0i} = B_i^{'}\beta_b + b_{0i}$ with its variance $\sigma_d^2 = \beta_b^{'}V_b\beta_b  + \sigma_b^2$, and the individual-level error term $\chi_{ij}^y = W_{ij}^{'}\beta_w + \epsilon_{ij}^y$ with its variance $\sigma_{\chi}^2 =  \beta_w^{'}V_w\beta_w  + \sigma_{\epsilon y}^2$. Note that $\Phi = \sigma_d^2 I_{n}+\sigma_{\chi}^2 J_{n}J_{n}^{^{\prime }}$. Now $d_{0i}$ is correlated with $T_{ij}$ such that $Cov(d_{0i}, T_{ij}) = \beta_b^{'}V_b\alpha_b$; and $\chi_{ij}^y$ is also correlated with $T_{ij}$ such that $cov(\chi_{ij}^y, T_{ij}) = \alpha_w^{'}V_w\beta_w$. These correlations lead to the violation of LMM assumptions and subsequently we cannot obtain consistent estimates of $\sigma_d^2$ or $\sigma_{\chi}^2$ or $\Phi $.

Hence, when unmeasured between- and within-cluster confounders exist, we assume positive and bounded constants $\sigma_{de}^2$ and $\sigma_{{\chi}e}^2$ such that $\widehat{\sigma}_d^2 \rightarrow_p \sigma_{de}^2$ and $\widehat{\sigma}_{\chi}^2 \rightarrow_p \sigma_{{\chi}e}^2$ as $m\rightarrow\infty$ for some estimators $\widehat{\sigma}_d^2$ and $\widehat{\sigma}_{\chi}^2$. Then, when $m\rightarrow\infty$, as shown in the Appendix, the bias of $\widehat{\beta}_L$ can be approximated as
\begin{eqnarray}
\widehat{\beta}_{L} - \beta  \rightarrow_p \frac{\alpha_b^{'}V_b\beta_b \frac{\sigma_{\chi e}^2}{\sigma_{\chi e}^2 + (n-1)\sigma_{de}^2} + \alpha_w^{'}V_w\beta_w }{(\sigma_a^2+\alpha_b^{'}V_b\alpha_b)\frac{\sigma_{\chi e}^2}{\sigma_{\chi e}^2 + (n-1)\sigma_{de}^2} + (\alpha_w^{'}V_w\alpha_w + \sigma_{\epsilon t}^2)}   \nonumber.
\end{eqnarray}
When both $m\rightarrow\infty$ and $n\rightarrow\infty$, $\widehat{\beta}_{L} - \beta \rightarrow_p \frac{ \alpha_w^{'}V_w\beta_w }{\alpha_w^{'}V_w\alpha_w + \sigma_{\epsilon t}^2}$. Note that the asymptotic bias here is the same as that of the FE estimator. This is consistent with the fact that the LMM estimator becomes the FE estimator when the cluster size increases. This is because the LMM estimator is a weighted combination of within and between cluster estimators, with weights depending on variance components and cluster size \cite{Fitz}.

When unmeasured within-cluster confounders but not between-cluster confounders exist, the bias of $\widehat{\beta}_{L}$ can be simplified as
\begin{eqnarray}
\widehat{\beta}_{L} - \beta  \rightarrow_p \frac{\alpha_w^{'}V_w\beta_w }{\sigma_a^2\frac{\sigma_{\chi e}^2}{\sigma_{\chi e}^2 + (n-1)\sigma_{de}^2} + (\alpha_w^{'}V_w\alpha_w + \sigma_{\epsilon t}^2)} \nonumber .
\end{eqnarray}
When both $m\rightarrow\infty$ and $n\rightarrow\infty$, $\widehat{\beta}_{L} - \beta \rightarrow_p \frac{\alpha_w^{'}V_w\beta_w }{ \alpha_w^{'}V_w\alpha_w + \sigma_{\epsilon t}^2}$, the same as above.

When unmeasured between-cluster but not within-cluster confounding exists, the bias of $\widehat{\beta}_{L}$ can be simplified as
\begin{eqnarray}
\widehat{\beta}_{L} - \beta  \rightarrow_p \frac{\alpha_b^{'}V_b\beta_b}{(\sigma_a^2+\alpha_b^{'}V_b\alpha_b) + \sigma_{\epsilon t}^2\frac{\sigma_{\chi e}^2 + (n-1)\sigma_{de}^2}{\sigma_{\chi e}^2}}  \nonumber .
\end{eqnarray}
When both $m\rightarrow\infty$ and $n\rightarrow\infty$, $\widehat{\beta}_{L} - \beta  \rightarrow_p 0$.

\section{Simulations}
We conduct simulations to examine: 1) how well the asymptotic bias formulae we derived approximate bias in finite samples for all four methods (i.e., as $m\rightarrow\infty$); 2) the bias patterns in the effect estimates from four methods when unmeasured between- and/or within-cluster confounders ($B$ and/or $W$) exist. We simulate $T_{ij}$ and $Y_{ij}$ using the true models ($\ref{truet1}$) and ($\ref{truey1}$) specified below:
\begin{eqnarray}
T_{ij} &=&  \alpha_{1c} + \alpha_{2c} C_{2ij} + \alpha_{3c} C_{3i} + \alpha_{1g} B_{1i} + \alpha_{1p} W_{1ij} + a_{0i} + \epsilon_{ij}^t, \nonumber \\
Y_{ij} &=& \beta_{1c} + \beta T_{ij} + \beta_{2c} C_{2ij} + \beta_{3c} C_{3i} + \beta_{1g} B_{1i} + \beta_{1p} W_{1ij} + b_{0i} + \epsilon_{ij}^y. \nonumber
\end{eqnarray}
Previously, in order to simplify the bias derivation, we assumed mean zeros for $W$, $B$, $T$ and $Y$ and no presence of $C$. We argued that the bias formulae should be the same with or without these assumptions. In simulations, we do not make these assumptions. The default parameter specifications are as follows: $m=200, n=20, \alpha_{1c}=18, \alpha_{2c} = -1, \alpha_{3c} = -1, \alpha_{1b} = 0.6, \alpha_{1w} = 0.6, \beta_{1c} = 3, \beta = 0.7, \beta_{2c} = 1, \beta_{3c} = 1, \beta_{1b} = 0.6, \beta_{1w} = 0.6$. We let $C_{2ij} \sim N(0, 1)$, $C_{3i} \sim N(11, 1)$, $B_{1i} \sim N(1, 1)$, $W_{1ij} \sim N(1, 1)$, $\epsilon_{ij}^t, \epsilon_{ij}^y \sim N(0, 1)$, $a_{0i} \sim N(0, 0.3^2)$, and $b_{0i} \sim N(0, 1)$. Various simulations with a wide range of parameter values were performed; all demonstrated similar patterns. Here we present the results when we vary one parameter while holding other parameters constant. For each set of parameter specifications, we simulate 5,000 and 1,000 data sets for $m=10$ and $m=200$ respectively since it requires more simulations to reach stable estimates when the number of clusters is small. We estimate the treatment dose effect $\beta$ using the OLS, LMM, FE, and IVA methods for each data set and then report the empirical bias of the estimates averaged over simulations. We also present analytical bias calculated directly from the asymptotic bias formulae we derived for the four estimators.

\subsection{When Unmeasured Within-cluster Confounders ($W$) Exist}
\begin{figure}
\centering
\includegraphics[width=12cm, height=17cm]{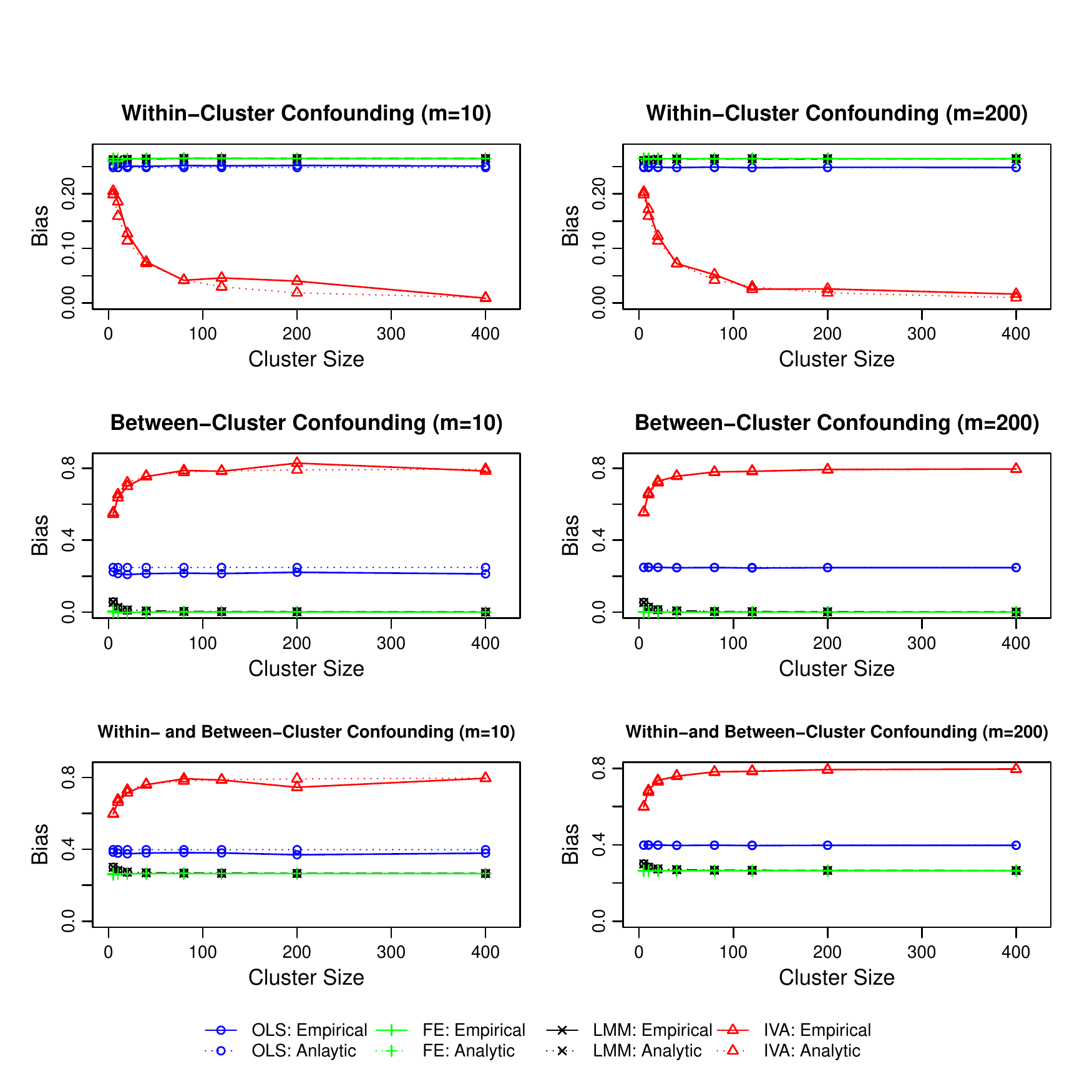}
\caption{Bias of OLS, FE, LMM and IVA estimates. Note that the FE and LMM lines are almost completely overlaid. The true treatment effect $\beta=0.7$. IVA: Preference-based IV regression; OLS: ordinary least squares regression; FE: fixed effect regression; LMM: linear mixed model. Unmeasured within-cluster confounding: $W$; Unmeasured between-cluster confounding: $B$; $m$: number of clusters; Empirical: averaged over simulations; Analytic: based on the asymptotic bias formulae. }
   \label{fig:2}
\end{figure}

\begin{figure}
\centering
\includegraphics[width=12cm, height=16.6cm]{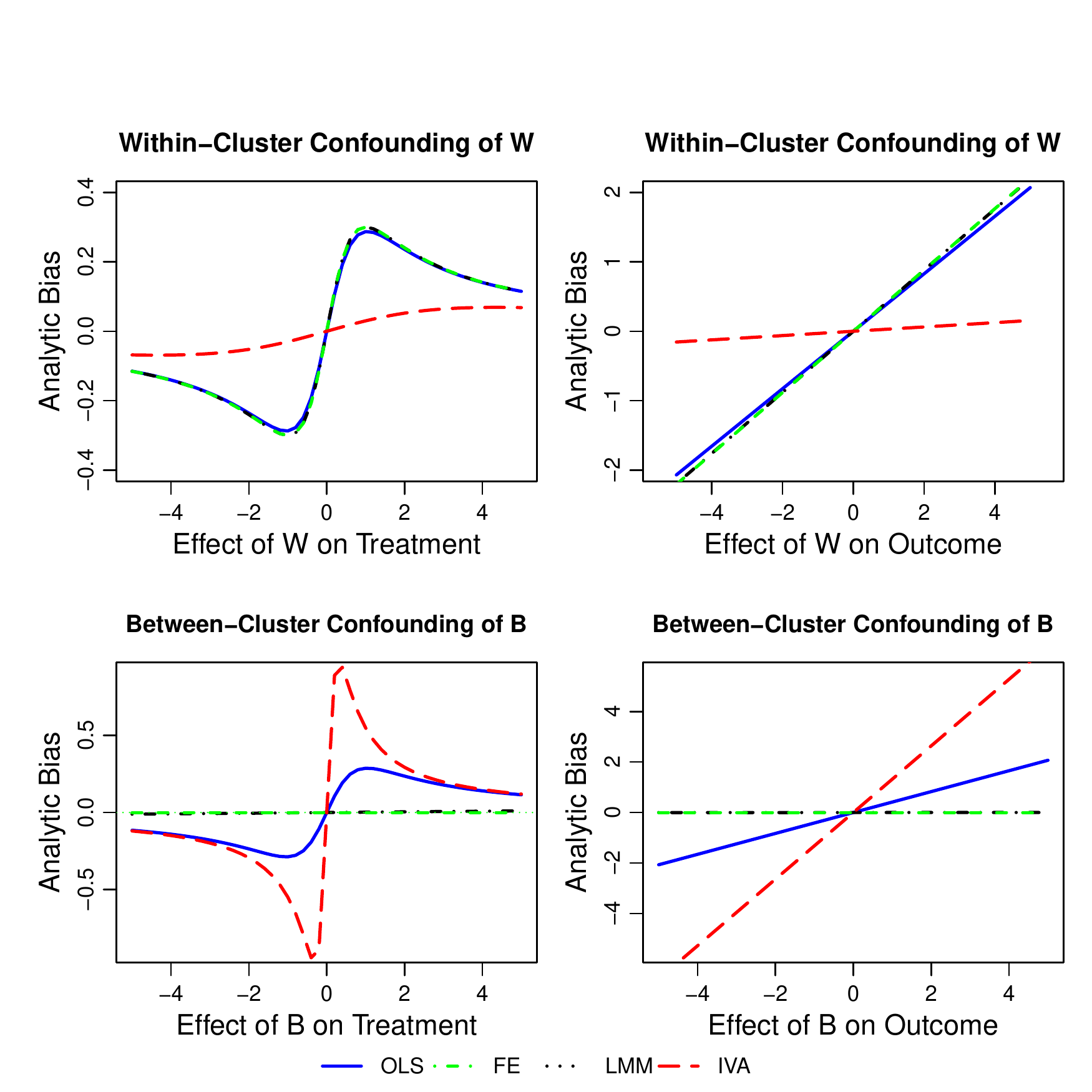}
  \caption{Bias of OLS, FE, LMM, and IVA estimates based on the bias formulae. Note that the FE and LMM lines are almost completely overlaid. OLS: ordinary least squares regression; FE: fixed effect regression; LMM: linear mixed model, IVA: Preference-based IV regression. The true treatment effect $\beta=0.7$. The cluster size $n = 200$. Within-cluster confounding: $W$ is unadjusted for; Between-cluster confounding: $B$ is unadjusted for; the effect of $W$ on treatment $T$, $\alpha_{1w}$; the effect of $W$ on outcome $Y$, $\beta_{1w}$; the effect of $B$ on $T$, $\alpha_{1b}$; the effect of $B$ on $Y$, $\beta_{1b}$. }
   \label{fig:3}
\end{figure}

When only unmeasured within-cluster confounders exist, we present the simulation results in the top panels of both Figures \ref{fig:2} and \ref{fig:3} to examine the bias of the estimates from the four methods. Figure \ref{fig:2} top panel shows the formulae approximate the finite-sample bias well for all methods, even when the number of clusters is very small (i.e., $m=10$). When $m$ is larger (i.e., $m=200$), the approximation is even better. Overall, when only unmeasured within-cluster confounders exist, IVA estimates are much less biased than OLS, FE, and LMM estimates. As the cluster size ($n$) approaches $1$, the bias of the IVA estimates approaches the OLS, FE and LMM estimates; as $n$ increases to around $400$, the bias of IVA estimates becomes negligible. In contrast, $n$ has little influence on the OLS, FE and LMM estimates. Figure \ref{fig:3} top panel shows the bias patterns of the four estimators based on the bias formulae for $n=200$ when either the effect of unmeasured within-cluster confounding on treatment dose ($\alpha_{1w}$) or on outcome ($\beta_{1w}$) vary. When $\alpha_{1w}$ departs further away from $0$, the magnitude of the bias for all estimates increases before decreasing. When the magnitude of $\beta_{1w}$ increases, the magnitude of bias of all estimates increases monotonically. Overall, the OLS, FE, and LMM estimates are similar, and the IVA estimates are much less biased. These simulation results are consistent with what the derived asymptotic formulae indicate.

\subsection{When Unmeasured Between-cluster Confounders ($B$) Exist}
When only unmeasured between-cluster confounders $B$ exist, Figure \ref{fig:2} middle panel demonstrates that the asymptotic bias formulae approximate the bias in finite samples very well for all four methods. The cluster size $n$ has the biggest impact on IVA estimates, and some impact on LMM estimates, but very little on OLS or FE estimates. As $n$ approaches $1$, the IVA and LMM estimates approach OLS estimates. In Figure \ref{fig:3} bottom panel, as the effect of $B$ on treatment $\alpha_{1b}$ departs further away from $0$, the biases of IVA and OLS estimates first increase and then decrease. As the effect of $B$ on outcome $\beta_{1b}$ becomes more different from $0$, the magnitudes of bias in both IVA and OLS estimates increase monotonically. Both Figures $\ref{fig:2}$ and $\ref{fig:3}$ demonstrate that when only unmeasured between-cluster confounders exist, IVA estimates have the largest bias in magnitudes, OLS estimates have some bias, and FE estimates have negligible bias. When $n$ is relatively small ($n<20$), LMM estimates have small finite-sample bias; as $n$ increases, their bias becomes negligible. Overall, the performance of FE and LMM estimators is similar, except when $n$ is small. These observed bias patterns are consistent with the derived bias formulae.

\subsection{When Both Unmeasured Within-Cluster and Between-Cluster Unmeasured Confounders ($W$ and $B$) Exist}
\begin{figure}
\centering
\includegraphics[width=12cm, height=16.5cm]{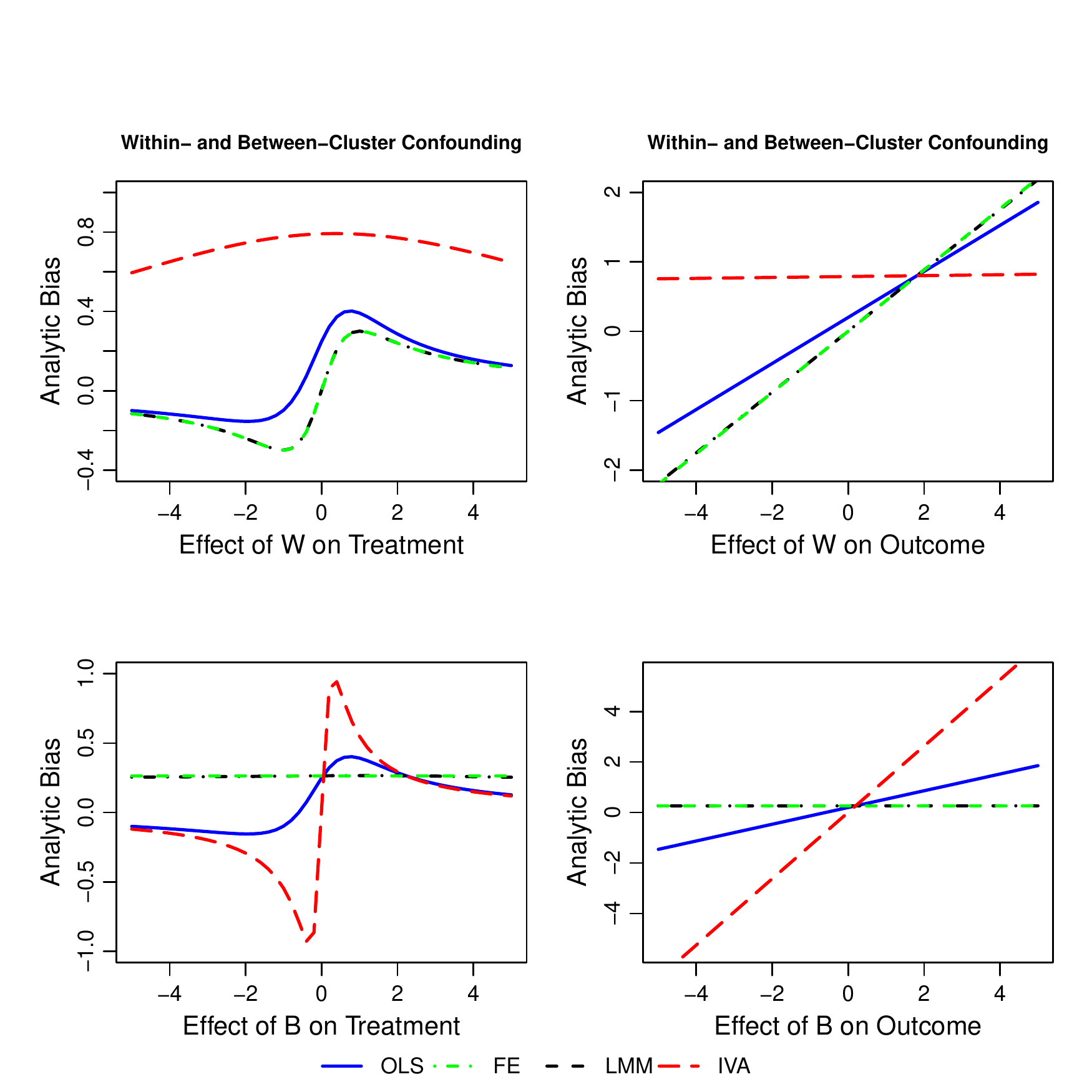}
  \caption{Bias of the OLS, FE, LMM, and IVA estimates based on the bias formulae. Note that the FE and LMM lines are almost completely overlaid. The true treatment effect $\beta=0.7$. The cluster size $n = 200$. OLS: ordinary least squares regression; FE: fixed effect regression; LMM: linear mixed model, IVA: Preference-based IV regression. Within-Cluster and Between-Cluster Confounding: neither $W$ nor $B$ are adjusted for. The effect of $W$ on the treatment $T$, $\alpha_{1w}$; the effect of $W$ on the outcome, $\beta_{1w}$; the effect of $B$ on the treatment, $\alpha_{1b}$; the effect of $B$ on the outcome, $\beta_{1b}$.}
   \label{fig:4}
\end{figure}

When both unmeasured within-cluster and between-cluster confounders exist, we summarize the simulation results in the bottom panel of Figure \ref{fig:2} and in Figure \ref{fig:4}. Figure \ref{fig:2} bottom panel shows that the asymptotic bias formulae approximate the finite-sample bias very well for all estimators, particularly when $m=200$. Here we assume the respective effects of $W$ and $B$ on $T$ and $Y$ are the same and set to $0.6$. We find that all estimates are biased with the IVA estimates having the largest biases, followed by OLS estimates and then LMM and FE estimates. The cluster size $n$ has the biggest impact on IVA estimates, some impact on LMM estimates and very little impact on OLS and FE estimates. As $n$ approaches 1, the bias of both IVA and LMM estimates approaches the bias of OLS estimates.

Figure \ref{fig:4} shows the bias patterns as the effects of $W$ and $B$ on $T$ and $Y$ increase based on the bias formulae. The magnitudes of the bias have complex relationships with the effects of $W$ and $B$ on $T$ ($\alpha_{1w}$ and $\alpha_{1b}$) for all four methods. On the other hand, the biases of all four estimators have linear relationships with both effects of $W$ and $B$ on $Y$ ($\beta_{1w}$ and $\beta_{1b}$). As $\beta_{1w}$ departs further away from $0$, the rate of increase in the magnitude of bias is the fastest among LMM and FE estimates, moderate among OLS estimates and negligible among IVA estimates. As $\beta_{1b}$ becomes more different from $0$, the rate of increase in the magnitude of bias is the fastest among IVA estimates, moderate among OLS estimates, and negligible among FE and LMM estimates. The IVA estimates are more immune to the impact of unmeasured within-cluster confounding while the FE and LMM estimates are more robust to the impact of unmeasured between-cluster confounding. When both $W$ and $B$ are present, which method is the least biased depends on the interplay between $W$ and $B$. When both $W$ and $B$ have the same effects on $T$ and $Y$, IVA estimates have largest bias because IVA estimates are much more sensitive to $B$ than what OLS, FE, and LMM estimates are to $W$. IV estimates can have smallest bias among these four estimators (results not shown), only when the effect of $W$ on $Y$ is overwhelmingly larger than the effect of $B$ on $Y$.

\section{A Case Study}
We compare these four methods in handling potential unmeasured between- and within-cluster confounders using DOPPS phase $3$ data (2005-2008) \cite{Pisoni} to estimate the effect of ESA on Hgb levels. The outcome of interest is the Hgb at the 14th month and the exposure of interest is the dose of ESA measured at one month prior to the outcome measure. These are chosen because it typically takes about four weeks for ESA to be fully effective in raising Hgb levels \cite{Eschbach}, and DOPPS started to collect monthly Hgb and ESA data in study phase 3. Our analytical sample includes $1,434$ dialysis patients in $67$ facilities. We consider the following covariates: patients characteristics (age, sex, race, years on dialysis, history of coronary artery disease, congestive heart failure, cancer, cerebrovascular disease, diabetes, gastrointestinal bleeding, peripheral vascular disease, hypertension, intravenous iron, psychiatric disorder, intravenous iron use) as well as facility quality indicators (i.e., percentage of patients using a central venous catheter for dialysis, serum albumin level $<3.5$ g/dL, serum phosphorus level $>5.5$ mg/dL, and single pool kt/V [a measure of dialysis dose] $<1.2$). We also consider patients' ESA responsiveness, an indicator of ESA dose required to raise or sustain Hgb level, and effectively captured by the combination of ESA dose at two months prior to the outcome Hgb and Hgb level at one month prior to the outcome Hgb (as demonstrated in our prior work \cite{LiY}). Prior work by others has often omitted these adjustments \cite{Madore}. We conduct three sets of analyses. The first set of analyses includes all variables, the second set excludes patients' ESA responsiveness variables and the third excludes ESA responsiveness variables and facility quality indicators. The results of three sets of analyses are summarized in Table \ref{Tab:2}. For the IVA, we choose the ESA dose preferences of dialysis facilities as the instrument. An instrument is called strong when the association between this instrument and the exposure of interest is strong \cite{Stock, Burgess}. The stronger the IV is, the smaller the finite sample bias is. The partial F statistic in the regression of the exposure on the instrument is usually used as a measure of the strength of an instrument \cite{Stock}. An instrument is considered weak if the F statistic is less than 10 \cite{Stock}. The F statistics for IVA in these three sets of analyses are 5, 6 and 6 respectively, indicating that a stronger instrument would be preferred \cite{Burgess, Stock, Staiger}.

In the first set of analyses adjusting for all variables, although the magnitudes of the estimated ESA effects differ across the four methods, all effects are positive and qualitatively consistent with the well-known fact that ESA increases Hgb levels \cite{Eschbach}. In the second set of analyses excluding ESA responsiveness variables, IVA continues to produce positive ESA effects, but the other three methods give falsely negative estimates. In the third set of analyses excluding both ESA responsiveness and facility quality indicators, the largest change in the estimated effect of ESA on Hgb, compared with the second set, is observed in the IVA method, while there is some change in the OLS estimate, a very small change in the LMM estimate, and no change in the FE estimate.

As demonstrated in our prior work \cite{LiY}, the ESA responsiveness variables are largely within-cluster confounders of the ESA-Hgb relationship. Hence, without adjusting for these variables, it is not surprising that the IVA estimate of ESA effect on Hgb does not change much and remains positive, given the robustness of IVA method to unmeasured within-cluster confounders. It is also not surprising that all other three methods are noticeably impacted and give falsely negative effect estimates, given their sensitivity towards unmeasured within-cluster confounders. When we further exclude facility quality indicators, which are between-cluster confounders, the biggest change occurs in the IVA estimate, with little or no change in the LMM and FE estimates. Again, this is consistent with our finding that LMM and FE are robust to unmeasured between-cluster confounding, but IVA is not. SAS code used in implementing OLS, FE, LMM and IVA is provided in github at https://github.com/yunliyunli/Unmeasured-Confounding.

\begin{table}[htbp]
\begin{center}
\begin{tabular}{lcccccc }
\hline
 & \multicolumn{2}{c}{All Vars} & \multicolumn{2}{c}{-- ESA Resp} & \multicolumn{2}{c}{ -- ESA Resp/Fac Vars } \\
\hline
Methods & Est  & $95\%$ CI & Est & $95\%$ CI & Est & $95\%$ CI  \\
\hline
IVA  & 0.039 & $(0.015, 0.063)$  & 0.047  & $(0.016, 0.078)$ & 0.10 & $(0.072, 0.127)$\\
OLS   &  0.018 & $(0.012, 0.024)$ & -0.003  & $(-0.010, 0.004)$ & 0.011 & $(0.004, 0.018)$ \\
FE   & 0.015 & $(0.009, 0.022)$  &-0.011 & $(-0.018, -0.005)$ & -0.011 & $(-0.018, -0.005)$ \\
LMM & 0.017 & $(0.010, 0.023)$ & -0.010 & $(-0.016, -0.003)$ & -0.008 & $(-0.015, -0.002)$ \\
\hline
\end{tabular}
\end{center}
\caption{Three sets of analyses to estimate the effect of ESA on Hgb. Note that ESA is known to increase Hgb levels and the effect of ESA on Hgb should be positive. All Vars: adjust for all covariates listed in text; -- ESA Resp: adjust for all but patient ESA responsiveness variables; -- ESA Resp / Fac Vars: adjust for all but patient ESA responsiveness variables and facility quality indicators. }
 \label{Tab:2}
\end{table}

\section{Discussion}

IVA is known for its advantages in combating unmeasured confounders. However, it is less known that certain commonly used alternative methods may be more robust to unmeasured confounding than IVA depending on the nature of the confounding. In this study, we focused on preference-based IVA (the most popular IV methods) with clustered data (the most common data structure for observational studies). We derived asymptotic bias for IVA and three alternative methods including OLS, FE, and LMM, when unmeasured between-cluster and/or within-cluster confounders exist. Simulations demonstrated that all bias formulae perform well in finite samples. We also examined the validity of the assumptions required for each method and the degree of bias when either type of unmeasured confounding exist. While almost none of the methods are valid when either type of unmeasured confounding exists, we found that some methods are more robust than others depending on the type of unmeasured confounding. When unmeasured within-cluster confounding is more probable, IVAs are least biased compared with other methods; on the other hand, when unmeasured between-cluster confounding is more probable, FE methods or LMMs are less biased than other methods. When both types of unmeasured confounding exist, it depends on which are the dominate unmeasured confounders. Our results provide guidance for researchers because the effect of each type of unmeasured confounding is specified directly with explicit and closed-form bias formulae.

Our analysis of bias gives insight into the robustness and assumption of each method in handling unmeasured confounders. In practice, it is more likely that both unmeasured between- and within-cluster confounders exist and that subsequently none of the methods have all their model assumptions met. When all methods give similar analysis results, it is likely that these results are robust towards unmeasured confounding. When effect estimates from these four methods differ substantially, as in our data analyses, it is likely that unmeasured confounders exist. We can examine which potential unmeasured within- or between-cluster confounders may exist with our substantive knowledge about the data and the research questions. If we believe strongly that unmeasured within-cluster confounding strongly dominates over unmeasured between-cluster confounding, IVA estimates should be closer to the truth than the other methods. On the other hand, if we believe that unmeasured between-cluster confounding dominates within-cluster confounding, we should believe the LMM or FE estimates are closer to the truth. IVA methods are more sensitive to unmeasured between-cluster confounding than FE or LMMs towards unmeasured within-cluster confounding. Hence, our research results are most useful when we have substantive knowledge about whether between- or within-cluster confounding is of major concern. For example, in our data, ESA responsiveness is known to be a dominating within-cluster (patient-level) confounder of the ESA-Hgb relationship and is of major concern if not adjusted for. In summary, our analysis of bias provide insights into the potential impact of unmeasured confounders and how we can interpret our analysis results accordingly. It also helps us become more aware of the strength and weakness of the four different methods. It gives us the opportunities to detect the presence of unmeasured confounding and subsequently adjust for more confounding. We can then translate our finding to broader audience better and more precisely.

Previous research has demonstrated that the effect of treatment dose itself can be partitioned into between- and within-cluster components \cite{Neuhuasb}. In this manuscript, we assume the between- and within-cluster treatment effects are the same. If the between-cluster level treatment effect is different from the within-cluster level treatment effect, it often implies there may be unmeasured confounders, selection bias or measurement errors \cite{Palta1}. All of these can be characterized as omitted covariate problems either at within-cluster or between-cluster levels. Without any omitted covariates, it often requires the between-cluster and within-cluster effects be equal for the effects to have causal interpretations \cite{Palta1}. It is worthwhile to point out that there is a connection between the treatment effect estimates from the preference-based IVA and the between-cluster treatment effect estimates as well as a connection between the treatment effect estimates from the fixed effect models and the within-cluster treatment effect estimates. These connections are worthy of further investigations.

Previous studies focused on the impact of between-cluster unmeasured confounding on mixed models \cite{Palta1, Chao, Palta2, Neuhuasb} or the validity of assumptions and finite sample bias for IVA \cite{LiY, Martens, Burgess}. To our knowledge, this study is the first to derive bias formulae for these four methods when unmeasured within-cluster and/or between-cluster confounding may exist, a common scenario in medical studies using clustered data sets. Our data generating models are LMMs, which are most commonly used in clustered data structures. We also assumed that unmeasured confounders are normally distributed and independent of other confounders in deriving the bias. In practice, it is likely that multiple unmeasured between- and within-cluster confounders exist, and many of them are correlated with other confounders. We simulated unmeasured confounding to represent the overall residual confounding after conditioning on other measured confounders and used the normal distribution to approximate the residual confounders, as often done \cite{Cessie}. Our results are general and form a foundation to perform sensitivity analyses. Our research focuses on continuous outcomes, continuous exposures and linear association between them. Our finding that between- and within-cluster unmeasured confounders have different impact on the bias of the effect estimators from these four types of models likely extends to other types of outcomes and exposures. It will be important to investigate the bias expression for other types of outcomes, exposures or non-linear associations between them, where the non-linearity could incur additional bias \cite{Lee}.

Our research quantifies the bias in the presence of unmeasured confounders. The results are also applicable to measurement error, selection bias, or selective drop-out issues because they can be cast into the same general framework of omitted variables. We assume the treatment effect is homogeneous, as done in most regression analyses. However, this assumption may not hold. When the effect is heterogeneous, IVA measures the treatment effect among compilers, that is, complier average treatment effect \cite{Imbens}, which may also explain why the IVA effect can be different from other model estimates. However, when there is no reason to believe the effect may be different for compliers, we can interpret treatment effect estimates in the context of unmeasured confounding.

In summary, we derived the bias formulae for IVA, LMM, FE, and OLS in the presence of unmeasured between- and within-cluster confounding. Our findings provide evidence to support future selection of methods to combat the dominant
types of unmeasured confounders, and facilitate the interpretation of statistical analysis results in the context of unmeasured confounding, and help detect the presence unmeasured confounders.

\begin{dci}
The authors declared no potential conflicts of interest with respect to the research, authorship,
and/or publication of this article.
\end{dci}

\begin{funding}
This work was supported in part by National Institutes of Health grants 5R01-DK070869 and UL1TR002240. The DOPPS is administered by Arbor Research Collaborative for Health and is supported by Amgen, Kyowa Hakko Kirin, AbbVie, Sanofi Renal, Baxter Healthcare, and Vifor Fresenius Medical Care Renal Pharma. Additional support for specific projects and countries is provided by Keryx Biopharmaceuticals, Merck Sharp \& Dohme, Proteon Therapeutics, Relypsa, and F Hoffmann-LaRoche; in Canada by Amgen, BHC Medical, Janssen, Takeda, and Kidney Foundation of Canada (for logistics support); in Germany by Hexal, DGfN, Shire, and WiNe Institute; and for PDOPPS in Japan by the Japanese Society for Peritoneal Dialysis. All support is provided without restrictions on publications.

\end{funding}

\begin{acks}
The authors thank Drs Min Zhang, Lu Wang and the reviewers for helpful suggestions, as well as Shauna Leighton for providing editorial assistance on this manuscript.
\end{acks}

\begin{spacing}{1}	

\end{spacing}


\section{Appendix}

\section{Appendix: Bias Derivation in the Presence of Both $W$ and $B$}

\subsection{Preference-based IVA Estimator}
In the absence of $C$, the two-stage generalized least squares estimator of $\beta$ in (\ref{2sgls}) can then be simplified to
\begin{eqnarray}
\widehat{\beta}_{I} &=& \left(\sum_{i=1}^m \widehat{T}_i^{'} \widehat{\Omega}^{-1} \widehat{T}_i \right)^{-1} \left(\sum_{i=1}^m \widehat{T}_i^{'} \widehat{\Omega}^{-1} Y_i \right). \nonumber
\end{eqnarray}

As $m\rightarrow \infty $, for some $\widehat{s}_{n,m}\rightarrow _{p}s_{n}$ introduced in Section 4.2.4, we note that
\begin{equation}
\widehat{\beta }_{I}-\beta =\frac{m^{-1}\sum_{i=1}^{m}\widehat{T}%
_{i}^{^{\prime }}(I_{n}-\widehat{s}_{n,m}J_{n}J_{n}^{^{\prime
}})(J_{n}b_{0i}+J_{n}B_{i}^{^{\prime }}\beta _{b}+W_{i}\beta _{w}+\epsilon
_{i}^{y})}{m^{-1}\sum_{i=1}^{m}\widehat{T}_{i}^{^{\prime }}(I_{n}-\widehat{s}%
_{n,m}J_{n}J_{n}^{^{\prime }})\widehat{T}_{i}}=\frac{A_{I}}{D_{I}}  \notag
\end{equation}%
under the true models (\ref{truet1}) and (\ref{truey1}) in the absence of $C$, where $\widehat{T}%
_{i}=Q_{n}T_{i}$ with $Q_{n}=J_{n}^{\prime }J_{n}/n$. For the numerator $%
A_{I}$, by applying the LLN for i.i.d. sequences as $m\rightarrow \infty $,
we have
\begin{eqnarray*}
 A_{I}&&\rightarrow_p lim_{m\rightarrow\infty} \frac{1}{m}\sum_{i=1}^{m}(J_{n}a_{0i}+J_{n}B_{i}^{^{\prime }}\alpha
_{b}+W_{i}\alpha _{w}+\epsilon _{i}^{t})^{^{\prime }}Q_{n}(I_{n}-\widehat{s}%
_{n,m}J_{n}J_{n}^{^{\prime }}) \\
&& \ \ \ \ \ \times (J_{n}b_{0i}+J_{n}B_{i}^{^{\prime }}\beta
_{b}+W_{i}\beta _{w}+\epsilon _{i}^{y}) \\
&&=\alpha _{b}^{^{\prime }}E(B_{i}J_{n}^{^{\prime
}}Q_{n}J_{n}B_{i}^{^{\prime }})\beta _{b}-s_{n}\alpha _{b}^{^{\prime
}}E(B_{i}J_{n}^{^{\prime }}Q_{n}J_{n}J_{n}^{^{\prime }}J_{n}B_{i}^{^{\prime
}})\beta _{b} \\
&&  \ \ \ \ +\alpha _{w}^{^{\prime }}E(W_{i}^{^{\prime }}Q_{n}W_{i})\beta
_{w}-s_{n}\alpha _{w}^{^{\prime }}E(W_{i}^{^{\prime
}}Q_{n}J_{n}J_{n}^{^{\prime }}W_{i})\beta _{w}  \\
&&=n\alpha _{b}^{^{\prime }}V_{b}\beta _{b}(1-ns_{n})+\alpha _{w}^{^{\prime
}}V_{w}\beta _{w}(1-ns_{n}) \text{.}
\end{eqnarray*}%
For the denominator $D_{I}$, since $Q_{n}J_{n}=J_{n}$ and $Q_{n}^{2}=Q_{n}$,
we similarly have%
\begin{eqnarray*}
D_{I} && \rightarrow_p lim_{m\rightarrow\infty} \frac{1}{m}\sum_{i=1}^{m}(J_{n}a_{0i}+J_{n}B_{i}^{^{\prime }}\alpha
_{b}+W_{i}\alpha _{w}+\epsilon _{i}^{t})^{^{\prime }}Q_{n} (I_{n}-\widehat{s}%
_{n,m}J_{n}J_{n}^{^{\prime }})Q_{n} \\
&& \ \ \ \ \ \times (J_{n}a_{0i}+J_{n}B_{i}^{^{\prime
}}\alpha _{b}+W_{i}\alpha _{w}+\epsilon _{i}^{t}) \\
&&=E(a_{0i}^{2})J_{n}^{^{\prime
}}Q_{n}J_{n}-s_{n}E(a_{0i}^{2})J_{n}^{^{\prime }}Q_{n}J_{n}J_{n}^{^{\prime
}}Q_{n}J_{n} +\alpha _{b}^{\prime }E(B_{i}J_{n}^{^{\prime }}Q_{n}J_{n}B_{i}^{^{\prime
}})\alpha _{b} \\ && \ \ \ -s_{n}\alpha _{b}^{\prime }E(B_{i}J_{n}^{^{\prime
}}Q_{n}J_{n}J_{n}^{^{\prime }}Q_{n}J_{n}B_{i}^{^{\prime }})\alpha _{b} +\alpha _{w}^{^{\prime }}E(W_{i}^{^{\prime }}Q_{n}W_{i})\alpha
_{w}  \\
&& \ \ \ -s_{n}\alpha _{w}^{^{\prime }}E(W_{i}^{^{\prime
}}Q_{n}J_{n}J_{n}^{^{\prime }}Q_{n}W_{i})\alpha _{w}  +E((\epsilon _{i}^{t})^{^{\prime }}Q_{n}\epsilon
_{i}^{t})-s_{n}E((\epsilon _{i}^{t})^{^{\prime }}Q_{n}J_{n}J_{n}^{^{\prime
}}Q_{n}\epsilon _{i}^{t})  \\
&&=n(\sigma _{a}^{2}+\alpha _{b}^{^{\prime }}V_{b}\alpha
_{b})(1-ns_{n})+(\alpha _{w}^{^{\prime }}V_{w}\alpha _{w}+\sigma _{\epsilon
t}^{2})(1-ns_{n})
\end{eqnarray*}%
as $m\rightarrow \infty $. Therefore, the asymptotic bias of $\widehat{\beta
}_{I}$ can be obtained as%
\begin{eqnarray*}
\widehat{\beta }_{I}-\beta && \rightarrow_p \frac{[n\alpha _{b}^{^{\prime }}V_{b}\beta
_{b}+\alpha _{w}^{^{\prime }}V_{w}\beta _{w}](1-ns_{n})}{[n(\sigma
_{a}^{2}+\alpha _{b}^{^{\prime }}V_{b}\alpha _{b})+(\alpha _{w}^{^{\prime
}}V_{w}\alpha _{w}+\sigma _{\epsilon t}^{2})](1-ns_{n})}  \\
&&=\frac{\alpha _{b}^{^{\prime }}V_{b}\beta _{b}+\alpha _{w}^{^{\prime
}}V_{w}\beta _{w}/n}{\sigma _{a}^{2}+\alpha _{b}^{^{\prime }}V_{b}\alpha
_{b}+(\alpha _{w}^{^{\prime }}V_{w}\alpha _{w}+\sigma _{\epsilon t}^{2})/n}%
\end{eqnarray*}%
as $m\rightarrow \infty $ for given $n$.

\subsection{OLS Estimator}
In the absence of $C$, the OLS estimator in ($\ref{olsest}$) can be simplified to:
\begin{eqnarray}
\widehat{\beta }_{O}=\left( \sum_{i=1}^{m}T_{i}^{\prime }T_{i}\right)
^{-1}\left( \sum_{i=1}^{m}T_{i}^{\prime }Y_{i}\right). \nonumber
\end{eqnarray}

We note that
\begin{equation}
\widehat{\beta }_{O}-\beta =\frac{m^{-1}\sum_{i=1}^{m}T_{i}^{^{\prime
}}(J_{n}b_{0i}+J_{n}B_{i}^{^{\prime }}\beta _{b}+W_{i}\beta _{w}+\epsilon
_{i}^{y})}{m^{-1}\sum_{i=1}^{m}T_{i}^{^{\prime }}T_{i}}=\frac{A_{O}}{D_{O}},
\notag
\end{equation}%
where as $m\rightarrow \infty $
\begin{eqnarray*}
A_{O}&& \rightarrow_p lim_{m\rightarrow\infty} \frac{1}{m}\sum_{i=1}^{m}(J_{n}a_{0i}+J_{n}B_{i}^{^{\prime }}\alpha
_{b}+W_{i}\alpha _{w}+\epsilon _{i}^{t})^{^{\prime
}}(J_{n}b_{0i}+J_{n}B_{i}^{^{\prime }}\beta _{b}+W_{i}\beta _{w}+\epsilon
_{i}^{y}) \\
&&=\alpha _{b}^{^{\prime }}E(B_{i}J_{n}^{^{\prime }}J_{n}B_{i}^{^{\prime
}})\beta _{b}+\alpha _{w}^{^{\prime }}E(W_{i}^{^{\prime }}W_{i})\beta
_{w}  \\
&&=n(\alpha _{b}^{^{\prime }}V_{b}\beta _{b}+\alpha _{w}^{^{\prime
}}V_{w}\beta _{w})
\end{eqnarray*}%
and
\begin{eqnarray*}
D_{O} && \rightarrow_p lim_{m\rightarrow\infty} \frac{1}{m}\sum_{i=1}^{m}(J_{n}a_{0i}+J_{n}B_{i}^{^{\prime }}\alpha
_{b}+W_{i}\alpha _{w}+\epsilon _{i}^{t})^{^{\prime
}}(J_{n}a_{0i}+J_{n}B_{i}^{^{\prime }}\alpha _{b}+W_{i}\alpha _{w}+\epsilon
_{i}^{t}) \\
&&=E(a_{0i}^{2})J_{n}^{^{\prime }}J_{n}+\alpha _{b}^{\prime
}E(B_{i}J_{n}^{^{\prime }}J_{n}B_{i}^{^{\prime }})\alpha _{b}+\alpha
_{w}^{^{\prime }}E(W_{i}^{^{\prime }}W_{i})\alpha _{w}+E((\epsilon
_{i}^{t})^{^{\prime }}\epsilon _{i}^{t}) \\
&&=n(\sigma _{a}^{2}+\alpha _{b}^{^{\prime }}V_{b}\alpha _{b}+\alpha
_{w}^{^{\prime }}V_{w}\alpha _{w}+\sigma _{\epsilon t}^{2}) \text{.}
\end{eqnarray*}%
Therefore, the asymptotic bias of $\widehat{\beta }_{O}$ can be obtained as
\begin{equation*}
\widehat{\beta }_{O}-\beta \rightarrow_p \frac{\alpha _{b}^{^{\prime }}V_{b}\beta
_{b}+\alpha _{w}^{^{\prime }}V_{w}\beta _{w}}{\sigma _{a}^{2}+\alpha
_{b}^{^{\prime }}V_{b}\alpha _{b}+\alpha _{w}^{^{\prime }}V_{w}\alpha
_{w}+\sigma _{\epsilon t}^{2}}
\end{equation*}%
as $m\rightarrow \infty $ for any $n$.

\subsection{Fixed-Effect Estimator}
In the absence of $C$, the FE estimator in (\ref{fe}) can be simplified to
\begin{eqnarray}
\hat{\beta}_{F} = \left(\sum_{i=1}^m T_i^{'} (I_n - Q_n) T_i \right)^{-1} \left(\sum_{i=1}^M T_i^{'} (I_n - Q_n) Y_i \right). \nonumber
\end{eqnarray}

We note that
\begin{equation}
\widehat{\beta }_{F}-\beta =\frac{m^{-1}\sum_{i=1}^{m}T_{i}^{^{\prime
}}(I_{n}-Q_{n})(J_{n}b_{0i}+J_{n}B_{i}^{^{\prime }}\beta _{b}+W_{i}\beta
_{w}+\epsilon _{i}^{y})}{m^{-1}\sum_{i=1}^{m}T_{i}^{^{\prime
}}(I_{n}-Q_{n})T_{i}}=\frac{A_{F}}{D_{F}}.  \notag
\end{equation}%
Since $(I_{n}-Q_{n})J_{n}=0$, cluster-level confounders (measured or
unmeasured) will not incur any bias in this case. Similarly above, as $%
m\rightarrow \infty $, we hence have%
\begin{eqnarray*}
lim_{m\rightarrow\infty}
A_{F} && \rightarrow_p \frac{1}{m}\sum_{i=1}^{m}(J_{n}a_{0i}+J_{n}B_{i}^{^{\prime }}\alpha
_{b}+W_{i}\alpha _{w}+\epsilon _{i}^{t})^{^{\prime
}}(I_{n}-Q_{n}) \\ && \ \ \ \ \ \times (J_{n}b_{0i}+J_{n}B_{i}^{^{\prime }}\beta _{b}+W_{i}\beta
_{w}+\epsilon _{i}^{y}) \\
&&=\frac{1}{m}\sum_{i=1}^{m}(W_{i}\alpha _{w}+\epsilon _{i}^{t})^{^{\prime
}}(I_{n}-Q_{n})(W_{i}\beta _{w}+\epsilon _{i}^{y}) \\
&&=\alpha _{w}^{^{\prime }}E(W_{i}^{^{\prime }}W_{i})\beta _{w}-\alpha
_{w}^{^{\prime }}E(W_{i}^{^{\prime }}Q_{n}W_{i})\beta _{w}  \\
&&=(n-1)\alpha _{w}^{^{\prime }}V_{w}\beta _{w}
\end{eqnarray*}%
and
\begin{eqnarray*}
D_{F} && \rightarrow_p lim_{m\rightarrow\infty} \frac{1}{m}\sum_{i=1}^{m}(J_{n}a_{0i}+J_{n}B_{i}^{^{\prime }}\alpha
_{b}+W_{i}\alpha _{w}+\epsilon _{i}^{t})^{^{\prime
}}(I_{n}-Q_{n}) \\ &&  \ \ \ \ \ \times (J_{n}a_{0i}+J_{n}B_{i}^{^{\prime }}\alpha _{b}+W_{i}\alpha
_{w}+\epsilon _{i}^{t}) \\
&&=\frac{1}{m}\sum_{i=1}^{m}(W_{i}\alpha _{w}+\epsilon _{i}^{t})^{^{\prime
}}(I_{n}-Q_{n})(W_{i}\alpha _{w}+\epsilon _{i}^{t}) \\
&&=\alpha _{w}^{^{\prime }}E(W_{i}^{^{\prime }}W_{i})\alpha _{w}-\alpha
_{w}^{^{\prime }}E(W_{i}^{^{\prime }}Q_{n}W_{i})\alpha _{w}+E((\epsilon
_{i}^{t})^{^{\prime }}\epsilon _{i}^{t})-E((\epsilon _{i}^{t})^{^{\prime
}}Q_{n}\epsilon _{i}^{t}) \\
&&=(n-1)(\alpha _{w}^{^{\prime }}V_{w}\alpha _{w}+\sigma _{\epsilon
t}^{2}) \text{.}
\end{eqnarray*}%
Therefore, as $m\rightarrow \infty $, the asymptotic bias of $\widehat{\beta }_{F}$ can be obtained as
\begin{equation*}
\widehat{\beta }_{F}-\beta \rightarrow_p \frac{\alpha _{w}^{^{\prime }}V_{w}\beta _{w}}{%
\alpha _{w}^{^{\prime }}V_{w}\alpha _{w}+\sigma _{\epsilon t}^{2}}.
\end{equation*}%

\subsection{LMM Estimator}

In the absence of $C$, the LMM estimator in ($\ref{lmest}$) can be simplified as
\begin{eqnarray}
\hat{\beta}_{L} = \left(\sum_{i=1}^m T_i^{'} \widehat{\Phi}^{-1} T_i \right)^{-1} \left(\sum_{i=1}^m T_i^{'} \widehat{\Phi}^{-1} Y_i \right). \nonumber
\end{eqnarray}

When unmeasured between- and within-cluster confounders exist, the cluster-level error term in ($\ref{lm}$) $d_{0i} = B_i^{'}\beta_b + b_{0i}$ with its variance $\sigma_d^2 = \beta_b^{'}V_b\beta_b  + \sigma_b^2$, and the individual-level error term $\chi_{ij}^y = W_{ij}^{'}\beta_w + \epsilon_{ij}^y$ with its variance $\sigma_{\chi}^2 =  \beta_w^{'}V_w\beta_w  + \sigma_{\epsilon y}^2$. Here $\Phi = \sigma_d^2I_n + \sigma_{\chi}^2J_n J_n^{'} = (\beta_b^{'}V_b\beta_b  + \sigma_b^2)I_n + (\beta_w^{'}V_w\beta_w  + \sigma_{\epsilon y}^2)J_n J_n^{'}$. We can obtain that $\Phi^{-1} = \frac{1}{\beta_w^{'}V_w\beta_w  + \sigma_{\epsilon y}^2}\{I_n - \frac{(\beta_b^{'}V_b\beta_b + \sigma_b^2)J_n J_n^{'}}{(\beta_w^{'}V_w\beta_w  + \sigma_{\epsilon y}^2) + n(\beta_w^{'}V_w\beta_w + \sigma_b^2)}\} = \frac{1}{\beta_w^{'}V_w\beta_w  + \sigma_{\epsilon y}^2} \{I_n - s_{n,m} J_nJ_n^{'}\}$ where $ s_{n,m} =\frac{\beta_b^{'}V_b\beta_b + \sigma_b^2}{(\beta_w^{'}V_w\beta_w  + \sigma_{\epsilon y}^2) + n(\beta_w^{'}V_w\beta_w + \sigma_b^2)}$. Now $d_{0i}$ is correlated with $T_{ij}$ such that $Cov(d_{0i}, T_{ij}) = \beta_b^{'}V_b\alpha_b$; and $\chi_{ij}^y$ is also correlated with $T_{ij}$ such that $cov(\chi_{ij}^y, T_{ij}) = \alpha_w^{'}V_w\beta_w$. These correlations lead to the violation of LMM assumptions and subsequently we cannot obtain consistent estimates of $\sigma
_{d}^{2}$ or $\sigma _{\chi }^{2}$. We instead assume some positive and
bounded constants $\sigma _{de}^{2}$ and $\sigma _{\chi e}^{2}$ such that $%
\widehat{\sigma }_{d}^{2}\rightarrow _{p}\sigma _{de}^{2}$ and $\widehat{%
\sigma }_{\chi }^{2}\rightarrow _{p}\sigma _{\chi e}^{2}$ as $m\rightarrow
\infty $ for some estimators $\widehat{\sigma }_{\chi }^{2}$ and $\widehat{%
\sigma }_{d}^{2}$. We then have $\widehat{s}_{n,m}\rightarrow _{p}s_{n}=%
\frac{\sigma _{de}^{2}}{\sigma _{\chi e}^{2}+n\sigma _{de}^{2}}$, which
satisfies $0<s_{n}<1$ for given $n$.

We note that
\begin{equation}
\widehat{\beta }_{L}-\beta =\frac{m^{-1}\sum_{i=1}^{m}T_{i}^{^{\prime
}}(I_{n}-\widehat{s}_{n,m}J_{n}J_{n}^{^{\prime
}})(J_{n}b_{0i}+J_{n}B_{i}^{^{\prime }}\beta _{b}+W_{i}\beta _{w}+\epsilon
_{i}^{y})}{m^{-1}\sum_{i=1}^{m}T_{i}^{^{\prime }}(I_{n}-\widehat{s}%
_{n,m}J_{n}J_{n}^{^{\prime }})T_{i}}=\frac{A_{L}}{D_{L}}  \notag
\end{equation}%
similarly as the IVA case. It follows that, as
$m\rightarrow \infty $, we have
\begin{eqnarray*}
A_{L} && \rightarrow_p lim_{m\rightarrow\infty} \frac{1}{m}\sum_{i=1}^{m}(J_{n}a_{0i}+J_{n}B_{i}^{^{\prime }}\alpha
_{b}+W_{i}\alpha _{w}+\epsilon _{i}^{t})^{^{\prime }}(I_{n}-\widehat{s}%
_{n,m}J_{n}J_{n}^{^{\prime }}) \\ && \ \ \ \ \ \times (J_{n}b_{0i}+J_{n}B_{i}^{^{\prime }}\beta
_{b}+W_{i}\beta _{w}+\epsilon _{i}^{y}) \\
&&=\alpha _{b}^{^{\prime }}E(B_{i}J_{n}^{^{\prime }}J_{n}B_{i}^{^{\prime
}})\beta _{b}-s_{n}\alpha _{b}^{^{\prime }}E(B_{i}J_{n}^{^{\prime
}}J_{n}J_{n}^{^{\prime }}J_{n}B_{i}^{^{\prime }})\beta _{b} \\
&& \ \ \ +\alpha _{w}^{^{\prime }}E(W_{i}^{^{\prime }}W_{i})\beta _{w}-s_{n}\alpha
_{w}^{^{\prime }}E(W_{i}^{^{\prime }}J_{n}J_{n}^{^{\prime }}W_{i})\beta
_{w}  \\
&&=n\alpha _{b}^{^{\prime }}V_{b}\beta _{b}(1-ns_{n})+n\alpha _{w}^{^{\prime
}}V_{w}\beta _{w}(1-s_{n})
\end{eqnarray*}%
and
\begin{eqnarray*}
D_{L} && \rightarrow_p lim_{m\rightarrow\infty} \frac{1}{m}\sum_{i=1}^{m}(J_{n}a_{0i}+J_{n}B_{i}^{^{\prime }}\alpha
_{b}+W_{i}\alpha _{w}+\epsilon _{i}^{t})^{^{\prime }}(I_{n}-\widehat{s}%
_{n,m}J_{n}J_{n}^{^{\prime }}) \\ &&  \ \ \ \ \ \times (J_{n}a_{0i}+J_{n}B_{i}^{^{\prime }}\alpha
_{b}+W_{i}\alpha _{w}+\epsilon _{i}^{t}) \\
&&=E(a_{0i}^{2})J_{n}^{^{\prime }}J_{n}-s_{n}E(a_{0i}^{2})J_{n}^{^{\prime
}}J_{n}J_{n}^{^{\prime }}J_{n} + \alpha _{b}^{\prime }E(B_{i}J_{n}^{^{\prime }}J_{n}B_{i}^{^{\prime
}})\alpha _{b} \\ && \ \ \ -s_{n}\alpha _{b}^{\prime }E(B_{i}J_{n}^{^{\prime
}}J_{n}J_{n}^{^{\prime }}J_{n}B_{i}^{^{\prime }})\alpha _{b} +\alpha _{w}^{^{\prime }}E(W_{i}^{^{\prime }}W_{i})\alpha _{w}  -s_{n}\alpha
_{w}^{^{\prime }}E(W_{i}^{^{\prime }}J_{n}J_{n}^{^{\prime }}W_{i})\alpha _{w}
\\
&&  \ \ \ +E((\epsilon _{i}^{t})^{^{\prime }}\epsilon _{i}^{t})-s_{n}E((\epsilon
_{i}^{t})^{^{\prime }}J_{n}J_{n}^{^{\prime }}\epsilon _{i}^{t}) \\
&&=n(\sigma _{a}^{2}+\alpha _{b}^{^{\prime }}V_{b}\alpha
_{b})(1-ns_{n})+n(\alpha _{w}^{^{\prime }}V_{w}\alpha _{w}+\sigma _{\epsilon
t}^{2})(1-s_{n}) \text{.}
\end{eqnarray*}%
Therefore, the asymptotic bias of $\widehat{\beta }_{L}$ can be obtained as%
\begin{eqnarray*}
\widehat{\beta }_{L}-\beta && \rightarrow_p \frac{\alpha _{b}^{^{\prime }}V_{b}\beta
_{b}(1-ns_{n})+\alpha _{w}^{^{\prime }}V_{w}\beta _{w}(1-s_{n})}{(\sigma
_{a}^{2}+\alpha _{b}^{^{\prime }}V_{b}\alpha _{b})(1-ns_{n})+(\alpha
_{w}^{^{\prime }}V_{w}\alpha _{w}+\sigma _{\epsilon t}^{2})(1-s_{n})}  \\
&& =  \frac{\alpha_b^{'}V_b\beta_b \frac{\sigma_{\chi e}^2}{\sigma_{\chi e}^2 + (n-1)\sigma_{de}^2} + \alpha_w^{'}V_w\beta_w }{(\sigma_a^2+\alpha_b^{'}V_b\alpha_b)\frac{\sigma_{\chi e}^2}{\sigma_{\chi e}^2 + (n-1)\sigma_{de}^2} + (\alpha_w^{'}V_w\alpha_w + \sigma_{\epsilon t}^2)}  \nonumber
\end{eqnarray*}%
as $m\rightarrow \infty $ for given $n$.

\end{document}